\documentclass[reprint,amsmath,amssymb,aps,nofootinbib]{revtex4-2}

\usepackage{graphicx}
\usepackage{bm}
\usepackage{siunitx}
\usepackage{booktabs}
\usepackage{microtype}
\usepackage{hyperref}
\usepackage{ulem}
\hypersetup{colorlinks=true,linkcolor=blue,citecolor=blue,urlcolor=blue}
\usepackage{float}
\usepackage{tikz}
\usepackage{tikz-feynman}
\usetikzlibrary{fit, backgrounds, shapes.geometric, positioning}
\usepackage{amsmath}
\usetikzlibrary{calc}
\usepackage{enumitem} 
\usepackage{makecell}
\usepackage{subfigure}

\begin{document}

\title{Inverse Excitation Hierarchy in Doubly-Heavy Tetraquarks within the Diquark Model}

\author{Maximilian Weber}
\email{weber@hken.phys.nagoya-u.ac.jp}
\affiliation{Department of Physics, Nagoya University, Nagoya 464-8602, Japan}
\author{Daiki Suenaga}
\email{suenaga@hken.phys.nagoya-u.ac.jp}
\affiliation{Kobayashi-Maskawa Institute for the Origin of Particles and the Universe, Nagoya University, Nagoya 464-8602, Japan}
\affiliation{Research Center for Nuclear Physics, Osaka University, Ibaraki 567-0048, Japan}
\author{Masayasu Harada}
\email{harada@hken.phys.nagoya-u.ac.jp}
\affiliation{Kobayashi-Maskawa Institute for the Origin of Particles and the Universe, Nagoya University, Nagoya 464-8602, Japan}
\affiliation{Department of Physics, Nagoya University, Nagoya 464-8602, Japan}
\affiliation{Advanced Science Research Center, Japan Atomic Energy Agency, Tokai 319-1195, Japan}

\date{\today}

\begin{abstract}
We investigate the $T_{cc}$ tetraquark, treating it as a bound state of a heavy diquark and a light antidiquark. Using the Silvestre-Brac potential and solving the Schrödinger equation via the Gaussian Expansion Method, we find that the excitation energy between the heavy diquark and light antidiquark is unexpectedly larger than that between the two light anti-quarks within the anti-diquark -- contrary to the naive expectation where the former is smaller than the latter. We trace this inversion of the mass hierarchy to the centrifugal force acting on the light degree of freedom. Applying the same framework to other systems ($T_{bb}, \Lambda_b, \Lambda_c$) yields qualitatively identical behavior, demonstrating the robustness of the mechanism. These results provide new insights into diquark dynamics and the mass structure of exotic hadrons.
\end{abstract}

\maketitle

\section{Introduction}
Hadrons, such as protons, neutrons, and mesons, are composite particles made of quarks bound by the strong force, described by Quantum Chromodynamics (QCD).  
Understanding hadron phenomena from the underlying theory of QCD remains one of the most challenging tasks in modern physics. The nonperturbative nature of QCD at low energies, e.g., the coupling constant growing large at long distances, causing confinement and chiral symmetry breaking, prevents the direct application of perturbation theory, and thus complicates the description of hadronic structures.

Hadrons can be categorized into baryons (three quarks) and mesons (a quark-antiquark pair), providing a framework to understand their properties and interactions.
Recent discoveries of several exotic hadrons, such as tetraquarks and pentaquarks (see, for reviews, e.g., Refs.~\cite{%
BRAMBILLA20201_The_XYZ_states__Brambilla, Chen_2023_Review_of_new_hadron_states__Chen, bicudo2022tetraquarkspentaquarkslatticeqcd})
exposes the limitations of conventional quark models, as their internal configurations can often be interpreted either as compact multiquark systems or loosely bound hadronic molecules, or others (e.g., as a triangle singularity).
Even for a simple baryon, the full QCD calculation is almost unsolvable, e.g., the quark-gluon interaction is not two-body, but many-body, and self-interaction is included. 
For multiquark states, this complexity increases combinatorially. Hence, one introduces effective degrees of freedom, such as diquarks, constituent quarks, or potential models, to make the problem solvable (see, e.g., Ref.~\cite{JAFFE20051, Diquarks_as_inspiration__Wilczek}).

The existence of exotic hadrons has long been anticipated in various theoretical frameworks \cite{PhysRevD.15.267_Multiquark_hadrons__Jaffe, Diquarks_as_inspiration__Wilczek, PhysRevLett.91.232003_Diquarks__Jaffe, PhysRevD.87.114511_LQCD_bb_tetraquark__Bicudo, PhysRevD.100.014503_LQCD_bb_tetraquark__Leskovec, PhysRevLett.131.161901_Doubly_Charmed_Tetraquarks_near_phy_point__Yan}.
Recent discoveries have highlighted the importance of studying exotic hadrons \cite{A_review_of_the_open_charm_and_open_bottom_systems__Chen}. 
In particular, the observation of the $T_{cc}^+$ tetraquark by the Large Hadron Collider beauty (LHCb) Collaboration at CERN \cite{cernTcc2021, LHCb2022Tcc} marked a milestone in hadron spectroscopy. This state, containing two charm quarks and two light antiquarks, exhibits an unusually long lifetime and an exceptionally narrow decay width of about 410~keV. 
Since its position is just below the $D^{*+}D^0$ threshold, many researchers consider its nature as a hadronic molecular state made from $D^*D$ (see, e.g., Ref.~\cite{Chen_2023_Review_of_new_hadron_states__Chen}, and references therein).
There are also many analyses using quark models (see, e.g., Refs.~\cite{PhysRevD.105.074021__Kim_Oka, lin2025massspectradoublycharmed__Lin, ORTEGA2023137918__Ortega, ANWAR2023138248__Anwar, He:2023ucd, Berwein_2024}).
Despite these advances, key aspects of doubly heavy tetraquarks remain unresolved. It is still unclear how quarks are spatially arranged within these states and how different configurations influence their binding and decay.  

In this paper, we study mass spectrum of the doubly heavy tetraquarks by means of the diquark ansatz developed in Ref.~\cite{PhysRevD.102.014004_Spectrum_of_singly_heavy_baryons__Kim}, where the chiral Effective Field Theory (EFT) was adopted to estimate masses of the scalar and pseudoscalar light diquarks with inputs from the lattice QCD~\cite{Bi:2015ifa}.
In contrast, in the present study, we employ the standard quark model Hamiltonian given in Ref.~\cite{Silvestre-Brac} with the Gaussian Expansion Method (GEM) \cite{HIYAMA2003223_GEM__Hiyama} to determine the diquark masses numerically.
Then, we reduce the tetraquark system by assuming that it consists of a light-diquark and a heavy-diquark. Each of these diquark systems is treated individually as a two-body system to obtain the corresponding diquark masses.

A novel result of the present study is that the $\rho$-mode excitation has a lower energy than expected. Thus, this mode is less energetic than the $\lambda$-mode. This kind of observation is a contradiction to our expectation where the $\rho$-mode should be located at a higher energy level, as demonstrated in the Harmonic Oscillator (HO) model (see, e.g., Refs.~\cite{Copley:1979wj, Nagahiro:2016nsx, GLOZMAN1996263_The_spectrum_of_the_nucleons__Glozman, PhysRevD.107.034031_Strong_decay_widths__Tecocoatzi}.).
Within the HO picture, excitations in different Jacobi coordinates are expected to be nearly degenerated, with the lighter reduced mass modes appearing slightly higher, which results in the $\rho$-mode being heavier than the $\lambda$-mode. 
In our present framework, such a {\it naive} mass hierarchy can be restored by tuning the mass of the light-diquark system. After certain threshold of $m_{ud}(0^-, 1^-, 2^-)=1.225$ GeV the $\rho$-mode will surpass the $\lambda$-mode energy in $T_{cc}$ system. In other words, by increasing the centrifugal energy inside the diquark, the naive hierarchy will be obtained.\\

This paper is organized as follows.
In Sec.~\ref{subsec:Diquark_Ansatz}, we introduce our picture to investigate $T_{cc}$ spectra based on the diquarks.
Next, in Sec.~\ref{sec:Hamiltonian_Construction} we discuss the employed Hamiltonian of an effective two-body problem within the diquark picture, as well as the determination of the range parameters needed for the GEM.
In Sec.~\ref{sec:results}, we present and discuss the numerical results for both diquark and tetraquark systems, including their mass spectra, wave functions, and root-mean-square radii.
Further insights into the internal structure and excitation behavior of the systems are analyzed in Sec.~\ref{sec:Tbb_and_Lamb}.
Finally, Sec.~\ref{sec:Conclusion} summarizes our findings and provides an outlook on possible extensions of this work.
Additional theoretical foundations are presented in the Appendix.

\section{
(Anti)Diquark in $T_{cc}$} \label{subsec:Diquark_Ansatz}

Since multi-quark systems are complicated in QCD, instead of treating every quark separately, it is convenient to focus on quarks forming correlated pairs.
Among them, diquarks, i.e., correlations of two quarks, play a significant role in describing hadrons~\cite{Ebert:1995fp, Diquarks_as_inspiration__Wilczek, PhysRevLett.91.232003_Diquarks__Jaffe, Harada:2019udr, Dmitrasinovic:2020wye, Santopinto:2004hw}. Motivated by this fact, in the present study, we adopt a light-antidiquark--heavy-diquark ($\overline{\text{LD}}$HD) picture to present $T_{cc}$ spectrum, where the light antidiquark ($\bar{u}\bar{d}$) and heavy diquark ($cc$) form the color-${\bm 3}$ and color-$\bar{\bm 3}$ states, respectively, as supported by their attractive forces (see Fig.~\ref{fig:Tcc_2pt_picture}). Here, the attractions in between those (anti)diquarks are described by a local potential, as constructed in Ref.~\cite{Silvestre-Brac}. We note that the relevant light antidiquark is a singlet in flavor and spin configurations, whereas the heavy diquark is spin-$1$, due to the Pauli principle. We also comment that other possible states for $T_{cc}$, such as molecular $D^{(*)}D^{(*)}$ states, are ignored, for which the diquark nature within $T_{cc}$ is examined in the simplest framework, in this exploratory work.
Although the $\bm6\otimes \bar{\bm6}$ configuration also satisfies the overall color-singlet condition, it is generally expected to play a subdominant role in the low-lying spectrum. In constituent quark models based on one-gluon exchange, the interaction in the color-$\bar{\bm 3}$ diquark channel is attractive, whereas the color-$\bm6$ channel is repulsive. Consequently, the $\bar{\bm 3}$ configuration is energetically favored and is usually regarded as the dominant component of compact tetraquark states. For this reason, the $\bm6\otimes \bar{\bm6}$ contribution is neglected in the present calculation. We note, however, that inclusion of this configuration could, in principle, lead to additional configuration mixing and may slightly modify the quantitative results.
For the $T_{cc}$ ground state, two possible $J^P = 1^+$ configurations can, in principle, be constructed within the dominant 
$\bm3\otimes \bar{\bm3}$ color configuration:
\begin{align}
    \Big[(cc)^{\bar{\bm3}}_{S=1}(\bar u\bar d)^{{\bm3}}_{S=0}\Big]_{J=1},
    \qquad
    \Big[(cc)^{\bar{\bm3}}_{S=1}(\bar u\bar d)^{{\bm3}}_{S=1}\Big]_{J=1}.
\end{align}
The first configuration contains the so-called ``good'' light antidiquark with isospin $I=0$, while the second involves the energetically less favored ``bad'' antidiquark configuration with $I=1$.
In the present study, however, we consider only the configuration with $S_{\bar q \bar q}=0$, which is expected to provide the contribution to the ground-state properties, as experiments already suggested~\cite{ParticleDataGroup}. Therefore, we restrict ourselves to assume that $T_{cc}$ is an isosinglet with $I=0$.

One of the most important properties to understand the structure of hadrons is chiral symmetry and its spontaneous breaking of light flavors. 
The former relates to states of opposite parity, the so-called chiral partners to each other. Its linear representation predicts that in the chiral-symmetry restored limit, the partners become degenerate, while its spontaneous breaking causes a mass difference between the partners. The light-(anti)diquark dynamics is also expected to be governed by such chiral dynamics, which motivated examinations of single heavy baryons based on chiral models~\cite{PhysRevD.70.031503_Chiral_doublers__Nowak, PhysRevD.85.014015_Lambdac_bound_states__Liu,PhysRevD.87.056007_Chiral_partner_structure__Harada_Ma,PhysRevD.97.114024_Analysis_of_Lambdac__Yohei_Harada,Kawakami:2019hpp,Kawakami:2020sxd, Suenaga:2023tcy, Suenaga:2024vwr, MA2015463_Doubly_heavy_baryons__Ma_Harada}.
To understand how chiral partners appear within our present picture of the $T_{cc}$ system, it is useful to separate orbital excitations into three types, as shown in Fig.~\ref{fig:Tcc_2pt_picture}. As described in this figure, the $\rho$-mode is an excitation inside the light-antidiquark itself, flipping its parity and giving it orbital angular momentum. Hence, it is expected that the chiral partners naturally line up with parity-flipped $\rho$-mode excitations of the light-diquark. The $\lambda$-mode, on the other hand, is an excitation between the heavy core and the light subsystem. We also classify $\rho_{cc}$-mode to refer to orbital excitations within the heavy-diquark itself.

\section{Hamiltonian Construction} \label{sec:Hamiltonian_Construction}
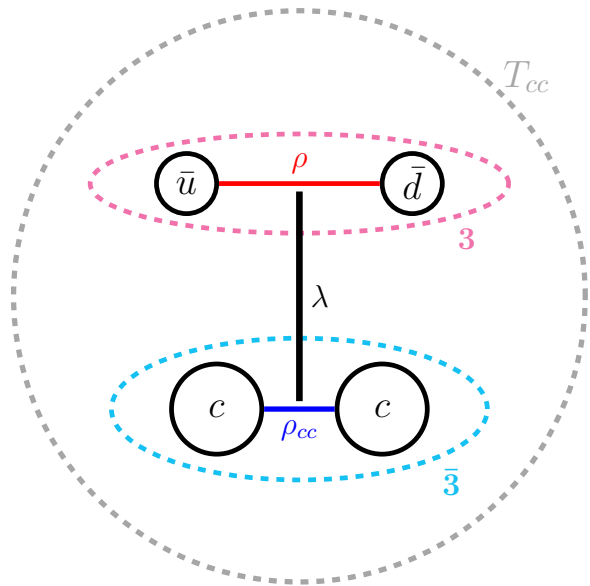
\begin{figure}
    \centering
    \begin{tikzpicture}
      \usetikzlibrary{fit, backgrounds, shapes.geometric, positioning}
    
      \tikzset{
        particle/.style = {circle, draw, line width=1.8pt, inner sep=0pt, font=\Large},
        light/.style    = {particle, minimum size=0.8cm},   
        heavy/.style    = {particle, minimum size=1.2cm}    
      }
    
      \node[light] (q1) at (0, 3) {$\bar{u}$};
      \node[light] (q2) at (3, 3) {$\bar{d}$};
    
      \node[heavy] (Q1) at (0.4, 0) {$c$};
      \node[heavy] (Q2) at (2.6, 0) {$c$};
    
      \draw[red, line width=1.8pt] (q1) -- (q2)
        node[midway, above, font=\large] {$\rho$};
      \draw[blue, line width=2.pt] (Q1) -- (Q2)
        node[midway, below, font=\large] {$\rho_{cc}$};
    
      \coordinate (topmid) at ($(q1)!0.5!(q2)$);
      \coordinate (botmid) at ($(Q1)!0.5!(Q2)$);
      \draw[line width=2.5pt, shorten <=3pt, shorten >=3pt]
        (topmid) -- (botmid)
        node[midway, right, font=\large] {$\lambda$};
    
      \begin{scope}[on background layer]
        \node[draw, ellipse, dashed, line width=1.8pt, magenta!70, fit=(q1)(q2), inner sep=1pt] (lightcluster) {};
        \node[draw, ellipse, dashed, line width=1.8pt, cyan!70, fit=(Q1)(Q2), inner sep=1pt] (heavycluster) {};
      \end{scope}

      \node[font=\large, anchor=north west, magenta!70]
        at (lightcluster.south east) {$\mathbf{3}$};
      \node[font=\large, anchor=north west, cyan!70]
        at (heavycluster.south east) {$\mathbf{\bar{3}}$};
    
      \coordinate (center) at ($(topmid)!0.5!(botmid)$);
      \draw[gray!70, dashed, line width=2pt] (center) circle [radius=3.8cm];
    
      \node[font=\Large\color{gray!70}, anchor=south west] at ($(center)+(2.6cm,2.6cm)$) {$T_{cc}$};
    
    \end{tikzpicture}

    \caption{Schematic picture of the $T_{cc}$ system (dashed gray ellipse) with a color-$\mathbf{3}$ light-antidiquark (dashed magenta ellipse) and a color-$\mathbf{\bar{3}}$ heavy-diquark (dashed cyan ellipse) subsystems, based on our $\overline{\text{LD}}$HD description. The subsystems consist either of two antiquarks (top), i.e., anti-up $\bar{u}$ and anti-down $\bar{d}$, or two charm-quarks $c$ (bottom).
    The interquark separations are $\rho$ (top and red solid line), $\rho_{cc}$ (bottom and blue solid line), and $\lambda$ (vertical and black solid line).}
    \label{fig:Tcc_2pt_picture}
\end{figure}

As explained in Sec.~\ref{subsec:Diquark_Ansatz}, we adopt the $\overline{\text{LD}}$HD ansatz in the present analysis, where we group the two light-antiquarks, $\bar{u}$ and $\bar{d}$, together, and the two heavy-quarks, $c$, respectively. We have visualized this idea in Fig.\@ \ref{fig:Tcc_2pt_picture}. To evaluate the corresponding diquark masses, we solve the nonrelativistic standard Diquark Model (DQM) Hamiltonian
\begin{align}
    H^{\rm diquark}_{ij} = m_{i} + m_{j} + \frac{\mathbf{p}_{ij}^2}{2\mu_{ij}}- \frac{3}{16} \sum_{a=1}^8 \left(\lambda_{i}^a \cdot \lambda_{j}^a \right) V_{ij}(\mathbf{r}_{ij})\,, \label{DQMHamiltonian}
\end{align}
with the quarks $(i,j)\in\{(c,c), (u,d)\}$, the quark masses $m_i$ with the corresponding relative momentum $\mathbf{p} = \frac{m_{i} \mathbf{p}_{j} - m_{j} \mathbf{p}_{i} }{m_{i} + m_{j}}$, and the diquark reduced mass $\mu_{ij} = \frac{m_{i} m_{j}}{m_{i} + m_{j}}$.
The $\mathbf{r}_{ij}$ is the interquark distance $\mathbf{r}_i - \mathbf{r}_j$, $\lambda_i^a$ are the $SU(3)$ Gell-Mann matrices for the color space with the color index $a$ acting on the quarks, and $V_{ij}(\mathbf{r}_{ij})$ is an arbitrary potential. The color term $-\frac{3}{16}\lambda_i^a \lambda_j^a$ is expected from one gluon exchange analysis. It has a vanishing effect between two well-separated color singlets but leads to a $1/2$ factor between the quark-quark interaction in a baryon and the quark-antiquark interaction in a meson: $V^{qq}(\mathbf{r}) = \frac{1}{2}V^{q\bar{q}}(\mathbf{r})$.
This ansatz allows us to reduce the degree of freedom of the $T_{cc}$ system to two, with a simpler version of the center-of-mass-frame $\overline{\text{LD}}$HD Hamiltonian as
\begin{align}
   H = m_{cc} + m_{\bar{u}\bar{d}} + \frac{\mathbf{p}^2}{2\mu}- \frac{3}{16} \sum_{a=1}^8 \left(\lambda_{cc}^a \cdot \lambda_{\bar{u}\bar{d}}^a \right) V_{cc,\bar{u}\bar{d}}(\mathbf{r}_{cc,\bar{u}\bar{d}})\,. \label{eq:2pt_Hamiltonian_CoM_frame}
\end{align}
Now the Gell-Mann matrices $\lambda_{i}^a$ are acting on the diquarks $i \in \{cc, \bar{u}\bar{d}\}$, and $\mathbf{r}_{cc, \bar{u}\bar{d}}$ describes the interdiquark distance.
The masses for the diquark and anti-diquark, $m_{cc}$ and $m_{\bar{u}\bar{d}}$, are used to describe the reduced mass
\begin{align}
    \mu = \frac{m_{cc} m_{\bar{u}\bar{d}}}{m_{cc} + m_{\bar{u}\bar{d}}}\,,
\end{align}
and the relative momentum
\begin{align}
    \mathbf{p} = \frac{m_{cc} \mathbf{p}_{\bar{u}\bar{d}} - m_{\bar{u}\bar{d}} \mathbf{p}_{cc} }{m_{cc} + m_{\bar{u}\bar{d}}}\,,
\end{align}
while here $\mathbf{p}_i$ are the momenta of each diquark $i\in\{cc, \bar{u}\bar{d}\}$.

Now, to describe the effective interaction $V_{ij}(\mathbf{r}_{ij})$ between quarks, we employ a QCD-inspired potential motivated by our underlying model. While phenomenological in nature, its structure reflects essential features derived from QCD. The short-range part originates from one-gluon exchange, which generates a Coulomb-like interaction. Relativistic corrections to this exchange term further introduce spin-dependent components, such as the contact spin-spin (also called hyperfine), spin-orbit, and tensor forces. Contrarily, the long-range part of the potential is associated with color confinement and is well supported by lattice QCD results, indicating that it rises linearly with the interquark separation.
Consequently, it is widely accepted that a realistic quark-quark potential should at least contain a $Coulomb + linear\ central$ part together with a $hyperfine\ interaction$ to account for spin-dependent mass splittings. In most hadronic systems, the effects of spin-orbit and tensor terms are comparatively weak and can be neglected at leading order. For this reason, we adopt a simplified potential of the form $Coulomb + linear + hyperfine$, which captures the dominant physical contributions while maintaining the model's analytical and computational tractability.
Furthermore, to ensure consistency and predictive power, the potential is required to reproduce the known spectra of mesons and baryons before being applied to multiquark systems. 

Among the available forms, the potential proposed by Bhaduri \textit{et al.\@} \cite{A_unified_potential_for_mesons_and_baryons__Bhaduri} provides an excellent balance between physical accuracy and simplicity, and has been successfully used in both two- and three-body calculations. We therefore employ this potential as the basis of our analysis. 
However, since an improved version has been developed by Silvestre-Brac \cite{Silvestre-Brac}, benefiting from refined parameters and additional experimental constraints, we use this updated potential. Its most general form is given by
\begin{align}
    \begin{split}
        V_{ij}(r) =& - \frac{\kappa \left( 1 - e^{-\frac{r}{r_c}}\right)}{r} + \lambda r^p - \Lambda \\ 
        &+ \frac{2\pi}{3m_im_j}\kappa'\left( 1 - e^{-\frac{r}{r_c}} \right) \frac{e^{-\frac{r^2}{r_0^2}}}{\pi^{\frac{3}{2}}r_0^3} \boldsymbol{\sigma}_i \boldsymbol{\sigma}_j\,,
    \end{split}
\end{align}
with
\begin{align}
    r_0(m_i, m_j) = A\left(\frac{2m_i m_j}{m_i + m_j}\right)^{-B}\,,
\end{align}
the relative distance $r = |\mathbf{r}_i - \mathbf{r}_j|$, and the Pauli matrices $\boldsymbol{\sigma}_i$ acting on the spin of state $i$. Reference~\cite{Silvestre-Brac} presented four different forms of this potential with belonging parameters, where each potential was suitable for the ground state as well as for the excited states. We have chosen \textit{AL1}, i.e., for \textit{All} mesons, \textit{Linear} confinement ($p=1$), and the parameter $r_c$ equal to zero (represented by the $1$).
The corresponding values are annotated in Tab.~\ref{tab:AL1_parameters}. However, a crucial property to note is that the potential is isospin independent.
\begin{table}[h]
    \centering
    \begin{tabular}{l | l}
        \toprule[1.2pt]
        \toprule[1.2pt]
        Parameter \hspace{0.15cm} & \hspace{0.15cm} Value \\
        \midrule[1.2pt]
        $p$             & \hspace{0.15cm} $1$ \\
        $r_c$           & \hspace{0.15cm} $0$ \\
        $\kappa$        & \hspace{0.15cm} $0.5069$ \\
        $\kappa'$       & \hspace{0.15cm} $1.8609$ \\
        $\lambda$       & \hspace{0.15cm} $0.1653~\text{GeV}^2$ \\
        $\Lambda$       & \hspace{0.15cm} $0.8321$ \\
        $A$             & \hspace{0.15cm} $1.6553~\text{GeV}^{B-1}$ \\
        $B$             & \hspace{0.15cm} $0.2204$ \\
        \midrule[0.8pt]
        $m_u, m_d$      & \hspace{0.15cm} $0.315~\text{GeV}$ \\
        $m_s$           & \hspace{0.15cm} $0.577~\text{GeV}$ \\
        $m_c$           & \hspace{0.15cm} $1.836~\text{GeV}$ \\
        $m_b$           & \hspace{0.15cm} $5.227~\text{GeV}$ \\
        \bottomrule[1.2pt]
        \bottomrule[1.2pt]
    \end{tabular}
    \caption{Model parameters and constituent quark masses for the AL1 Silvestre-Brac potential~\cite{Silvestre-Brac}.}
    \label{tab:AL1_parameters}
\end{table}

To solve now the two-body problem, we employ the Gaussian Expansion Method (GEM) (see App.~\ref{app:GEM}) and use Eqs.\@ \eqref{eq:generalized_EV_problem}-\eqref{eq:GEM_potential_matrix_form} together with the AL1 Silvestre-Brac potential given as
\begin{align}
    V_{\bar{u}\bar{d}, cc}^{\text{AL1}}(r) =& - \frac{\kappa}{r} + \lambda r - \Lambda + \frac{2\pi}{3m_{\bar{u}\bar{d}} m_{cc}}\kappa' \frac{e^{-\frac{r^2}{r_0^2}}}{\pi^{\frac{3}{2}}r_0^3} \boldsymbol{\sigma}_{\bar{u}\bar{d}} \boldsymbol{\sigma}_{cc} \,. \label{SBPotential}
\end{align}
This assumption is motivated by the color structure of the system. A diquark in the color-$\bar{\bm 3}$ representation transforms under color $SU(3)$ in the same way as an antiquark, while an antidiquark in the color-$\bm 3$ representation behaves analogously to a quark. Consequently, the diquark–antidiquark pair forms a color-singlet configuration with the same color structure as an ordinary quark-antiquark meson. Under this approximation, it is therefore reasonable to employ an effective interaction potential similar to that used in conventional quark–antiquark systems. We note, however, that the diquark is an extended composite object rather than a pointlike constituent, and additional structure effects may modify the interaction beyond the present approximation.

\begin{figure}
    \centering
    \includegraphics[width=\linewidth]{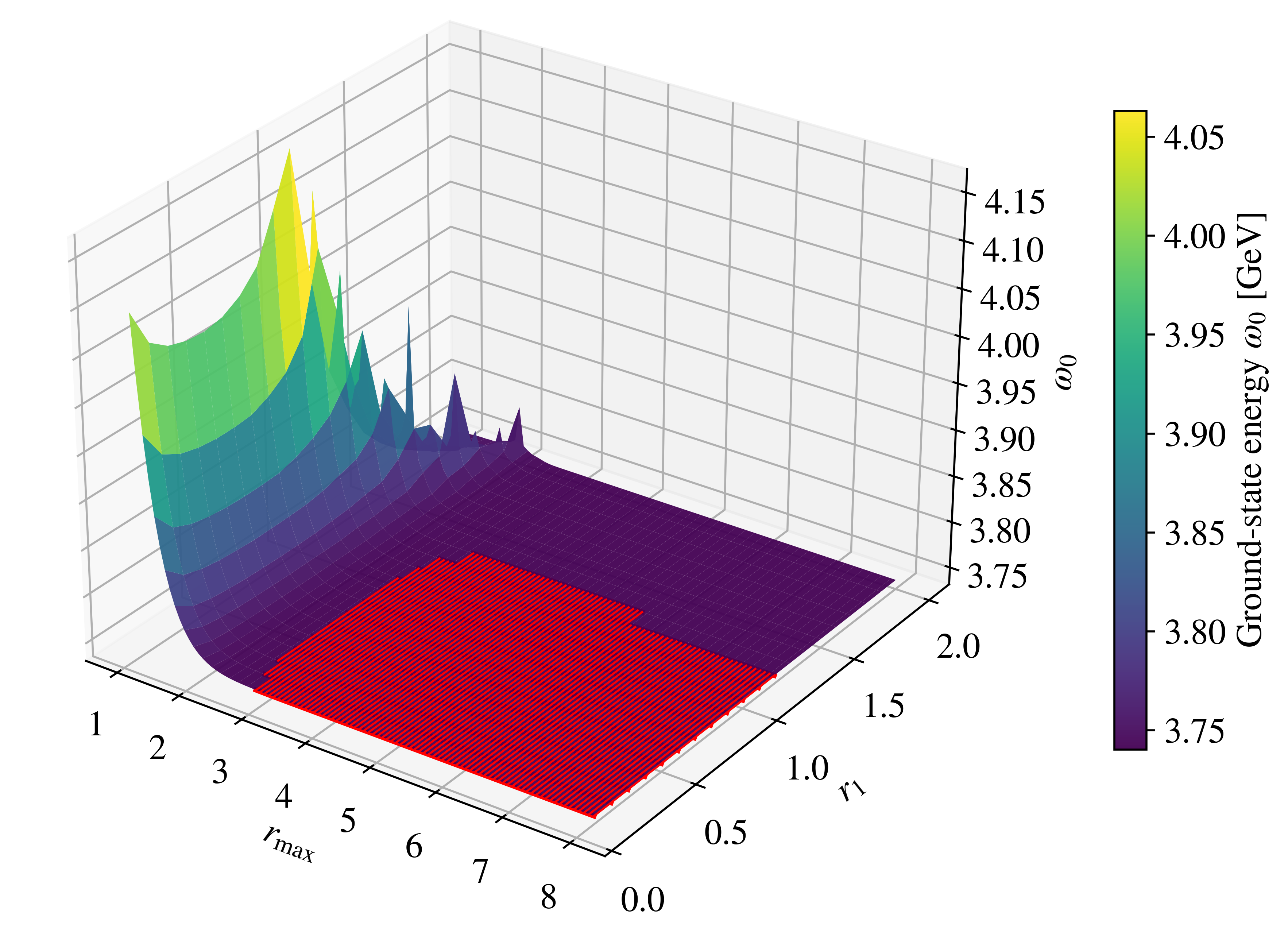}
    \caption{Energy landscape of the $T_{cc}$ system for $n_\text{max}= 20$ as a function of the nonlinear range parameters $(r_1, r_\text{max})$. The red area corresponds to the most stable and lowest-energy configurations where the energy difference was less than $1$~MeV. This area contains $956$ possible combinations and has an energy value of $3.740\pm 0.001$~GeV, with $m_{\bar{u}\bar{d}}=0.666\pm0.001$~GeV and $m_{cc}=3.500\pm0.001$~GeV.}
    \label{fig:Heatmap_nmax20_GS_Tcc}
\end{figure}

\begin{figure}
    \centering
    \includegraphics[width=\linewidth]{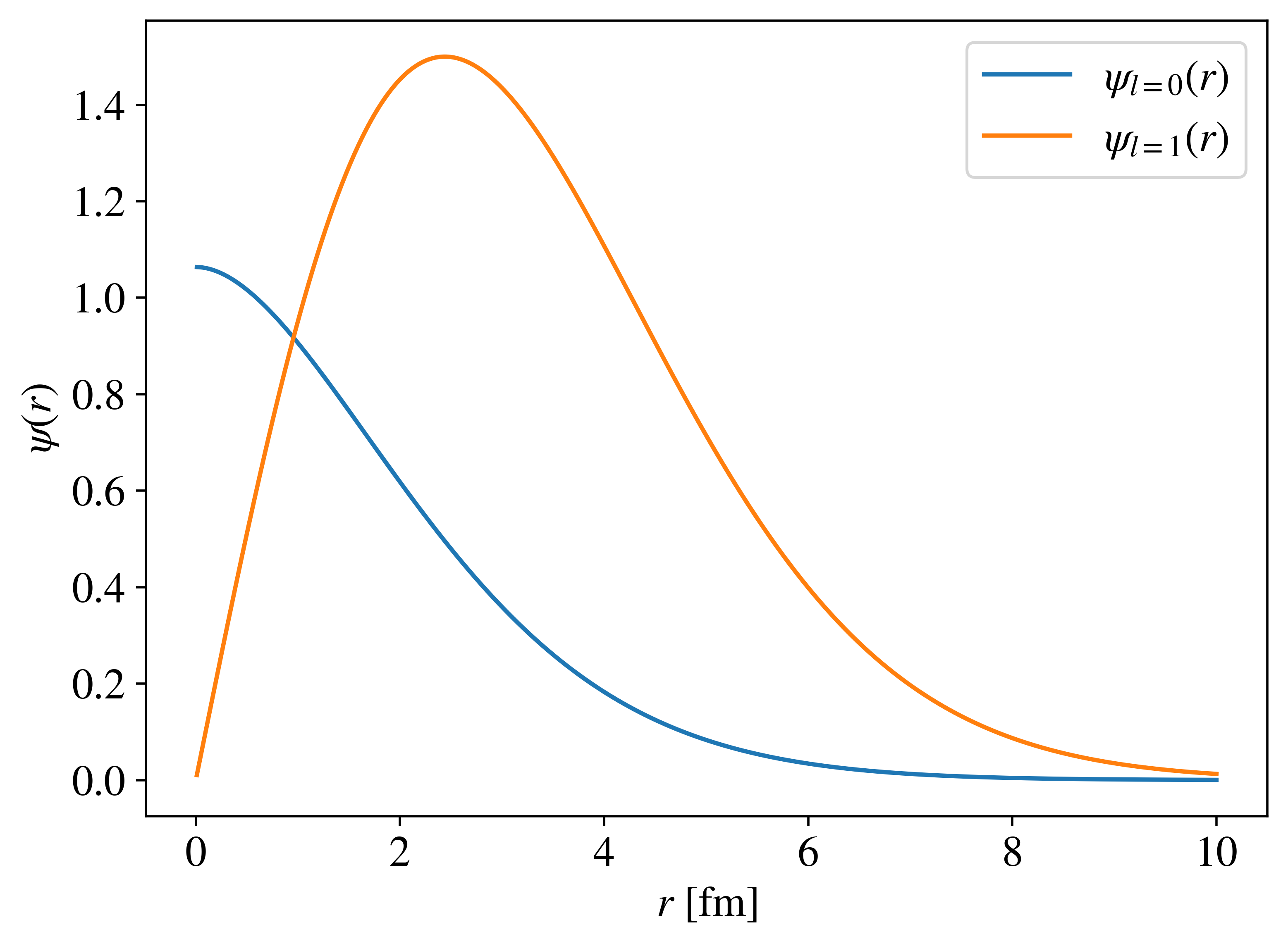}
    \caption{Radial wave function of the $T_{cc}$ ground (blue) and excited state (orange) obtained with the optimized range parameters $\{n_\text{max}=20, r_1 = 0.1\, \text{fm}, r_\text{max}=6\, \text{fm}\}$. The smooth and localized shape confirms the stability and convergence of the basis. Note that although the excited wave function extends beyond $6$~fm and approaches zero around $10$~fm, the obtained energies remain unchanged under the choice of $r_\text{max}$. In addition, choosing $6$~fm helps to reduce computational time.}
    \label{fig:Wavfunction_nmax20_GS_Tcc}
\end{figure}

First, the diquark masses are determined within the two-body framework and subsequently employed in the $T_{cc}$ calculation. Both diquark masses are obtained using the same approach, differing only in the corresponding quark-mass parameters and the spin-spin interaction.
In order to optimize the nonlinear range parameters in GEM, as pointed out in numerous studies \cite{10.1093/ptep/pts015_GEM__Hiyama, GEM__Hiyama, PhysRevA.85.022502__Hiyama}, a trial-and-error procedure combined with experience and systematic knowledge is performed to effectively describe the size and shape of the interaction, as well as the spatial extension of the system. In the present work, we refine this procedure further by scanning a wide range of range-parameter combinations $(r_1, r_\text{max})$ for each basis truncation $n_\text{max}$ (see App.~\ref{app:GEM} for the definition of these parameters). For every configuration, we calculate all corresponding energy eigenvalues and examined the stability of both the ground and excited states. Additionally, we inspect the resulting wave functions to ensure that physically stable and convergent solutions were obtained. This systematic cross-check allows us to determine robust and well-optimized range parameters tailored to the current system: 
\begin{align}
    \{n_\text{max}=20, r_1 = 0.1\, \text{fm}, r_\text{max}=6\, \text{fm}\}\,.
\end{align}
In Fig.\@ \ref{fig:Heatmap_nmax20_GS_Tcc} and Fig.\@ \ref{fig:Wavfunction_nmax20_GS_Tcc}, we illustrate this analysis for our chosen parameters for the case of $T_{cc}$. Note that we do this analysis not only for the tetraquark masses but also for the individual diquark masses.

\section{Results} \label{sec:results}
    \begin{table}
    \centering
    \begin{tabular}{l|c|c}
        \toprule[1.2pt]
        \toprule[1.2pt]
        (GeV) & Diquark Model & Chiral EFT Model \\
        \midrule[1.2pt]
         $m_{ud}(0^+)$ & $0.666$ & $0.725$ \cite{PhysRevD.102.014004_Spectrum_of_singly_heavy_baryons__Kim}\\
         $m_{ud}(0^-, 1^-, 2^-)$ & $1.121$ & $1.484$ \cite{PhysRevD.102.014004_Spectrum_of_singly_heavy_baryons__Kim}\\
         $m_{cc}(1^+)$ & $3.500$ & / \\
         $m_{cc}(1^-)$ & $3.701$ & / \\
         \midrule[0.8pt]
         $T_{cc}(1^+)$ & $3.740$ & $(3.775)$ \\ 
         $T_{cc}(0^-,1^-,2^-)_\lambda$ & $4.135$ & $(4.167)$ \\  
         \makecell[l]{$T_{cc}(1^-;0^-,1^-,2^-;$\\ $\phantom{T_{cc}(}1^-,2^-,3^-)_\rho$} & $4.054$ & $(4.346)$ \\ 
         $T_{cc}(1^-)_{\rho_{cc}}$ & $3.939$ & $(3.974)$ \\  
         \bottomrule[1.2pt]
         \bottomrule[1.2pt]
    \end{tabular}
    \caption{Comparison of the calculated diquark masses $m_{ud}$ and $m_{cc}$ used as input for the $T_{cc}$ system with results from the chiral EFT framework of Ref.~\cite{PhysRevD.102.014004_Spectrum_of_singly_heavy_baryons__Kim}. 
    Values in parentheses denote $T_{cc}$ masses obtained using the chiral model parameters for $m_{ud}$ but our predicted $m_{cc}$. Our approach yields slightly lower light-diquark masses and a reduced excitation gap by about $200$~MeV, while maintaining a similar overall spectral hierarchy.}
    \label{tab:Tcc_mud_mcc_comparison}
\end{table}

To begin our analysis with the $T_{cc}$ for isospin $I=0$, we first calculate the masses of the individual diquarks, $(ud)$ and $(cc)$, which serve as essential input parameters for the subsequent $T_{cc}$ calculation.
By solving the Schr\"{o}dinger equation with the DQM Hamiltonian~(\ref{DQMHamiltonian}), we predict the ($ud$) and ($cc$) diquark masses to be $0.666$~GeV and $3.500$~GeV, respectively,  which agree with the masses calculated in Ref.~\cite{lin2025massspectradoublycharmed__Lin}. Then, the $T_{cc}$ mass spectrum is obtained based on the $\overline{\rm LD}$HD ansatz with the potential~(\ref{SBPotential}). The results are summarized in Tab.~\ref{tab:Tcc_mud_mcc_comparison}.
We note that $T_{cc}$ ground state and $\lambda$-mode masses predicted in Ref.~\cite{lin2025massspectradoublycharmed__Lin} are $\sim100$~MeV higher than our DQM ones due to their adjustment of the $\kappa$ and $\lambda$ parameters.
Nevertheless, the difference between these two states is similar to ours, i.e., $\sim400$~MeV, verifying the validity of our model.
In Tab.~\ref{tab:Tcc_mud_mcc_comparison}, we also present the diquark masses estimated from a chiral EFT framework in Ref.~\cite{PhysRevD.102.014004_Spectrum_of_singly_heavy_baryons__Kim} for comparison, in order to assess possible systematic differences between the models. For their predictions, they used $m_{ud}(0^+) = 0.725$~GeV from lattice data~\cite{Bi:2015ifa} as input and calculated the excited masses based on the mass relations derived from their chiral EFT Lagrangian. 

We find that the mass of the light $(ud)$ diquark is slightly lower in our approach, whereas its excitation energy between the ground and first excited state is approximately $200$ MeV smaller. The deviation in absolute energies can be attributed to the different modeling methods, but the reduced excitation gap is particularly noteworthy.
Since no independent reference value for the $(cc)$ diquark mass is available, we use our predicted value as input for the $T_{cc}$ calculation in the alternative framework. 

\begin{figure}
    \centering
    \includegraphics[width=\linewidth]{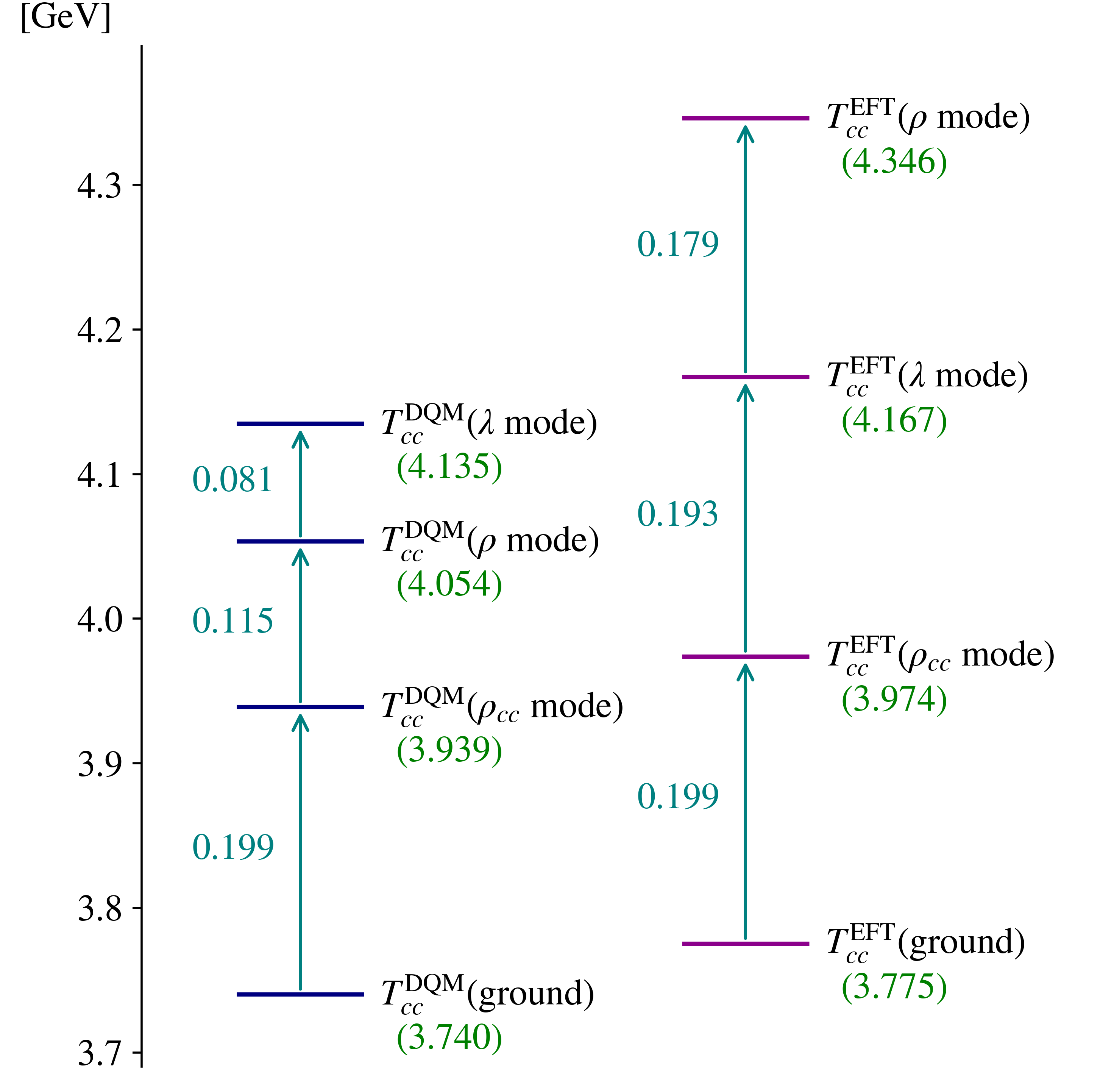}
    \caption{Comparison of the predicted $T_{cc}$ spectrum obtained in the DQM (blue solid lines) with that from the chiral EFT framework \cite{PhysRevD.102.014004_Spectrum_of_singly_heavy_baryons__Kim} (violet solid lines). The energy levels of corresponding states are shown together with their excitation energies relative to the ground state. While most states appear close in energy between the two approaches, the $\rho$-mode excitation lies approximately $300$ MeV lower in our model, indicating a significant sensitivity to the treatment of short-range diquark correlations.}
    \label{fig:spectrum_Tcc}
\end{figure}

The obtained $T_{cc}$ mass spectrum is visualized in Fig.~\ref{fig:spectrum_Tcc}. Here, $T_{cc}^{\rm DQM}$ corresponds to our results within the DQM, whereas $T_{cc}^{\rm EFT}$ are the chiral EFT ones \cite{PhysRevD.102.014004_Spectrum_of_singly_heavy_baryons__Kim} depicted as a reference. 
We note that our predicted values for the $T_{cc}$ ground state are lower than the $3875$~MeV~\cite{ParticleDataGroup} measured by experiments. This is due to our model approximation, in which we neglect concrete coupled-channel dynamics, possible pair creation, three-body forces, and more. However, in this study, we are not interested in the correct, real values but rather in the ordering of the different modes and possible anomalies, i.e., that the $\rho$-mode may be lighter than the $\lambda$-mode.\footnote{The same argumentation applies later to the different investigated cases.}
The $T_{cc}^{\rm DQM}$ spectrum shows that the $\rho$-mode excitation lies between the $\rho_{cc}$- and $\lambda$-modes, which contrasts with naive HO intuition. Meanwhile, $T_{cc}^{\rm EFT}$ obeys the naive hierarchy where the $\rho$-mode excitation energy is greater than the $\lambda$-mode one. It should be noted that the mass spectrum predicted by the three-body light-antidiquark--heavy-quark ansatz with the Yoshida potential \cite{PhysRevD.92.114029_Spectrum_of_heavy_baryons__Yoshida} and the chiral EFT also leads to the naive hierarchy of Ref.~\cite{PhysRevD.102.014004_Spectrum_of_singly_heavy_baryons__Kim} and Ref.~\cite{PhysRevD.105.074021__Kim_Oka}, respectively.

\begin{figure}
    \centering
    \includegraphics[width=0.9\linewidth]{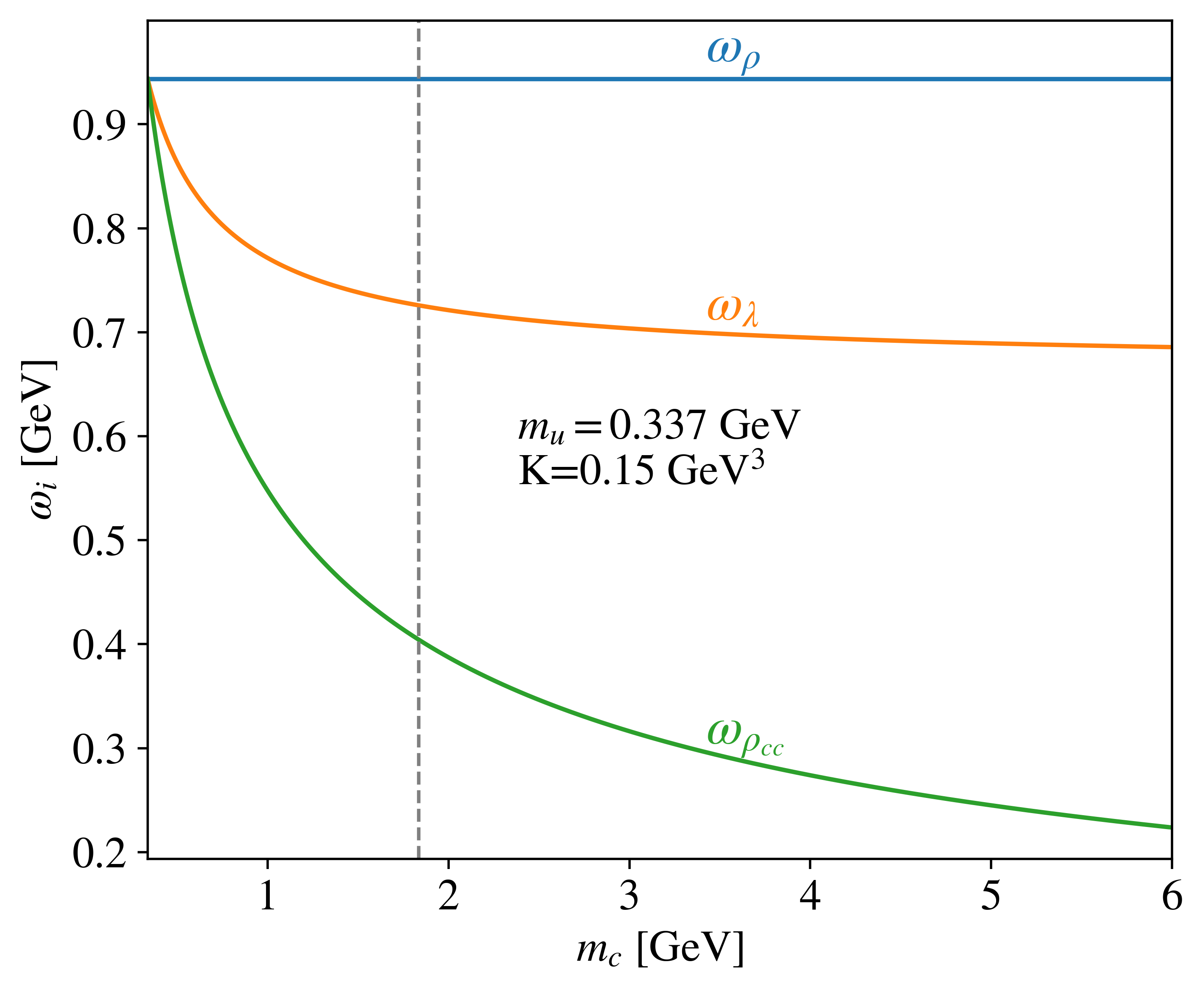}
    \caption{Heavy-quark mass ($m_c$) dependence of excited energies based on the HO model (see Eq.\@ \eqref{eq:HO_Jacobian_coord})
    for the $T_{cc}$ system. The $\rho$-mode (blue solid line), $\lambda$-mode (orange solid line), and the $\rho_{cc}$-mode (green solid line) are calculated by Eq.\@ \eqref{eq:oscillator_frequcies_HO_Jacobian_coord}. The vertical gray line corresponds to the used charm mass $m_c=1.836$~GeV \cite{Silvestre-Brac}. The initial parameters, $m_u$ and $K$, are given by the AL1 potential (see.\@ Tab.~\ref{tab:AL1_parameters}) and arbitrarily chosen, respectively.}
    
    \label{fig:HO_energy_dep}
\end{figure}

In the HO model (see App.~\ref{app:HO}), the oscillator frequencies are evaluated by
\begin{align}
    \omega_\rho = \sqrt{\frac{K}{\mu_\rho}}\,, \hspace{0.3cm} \omega_\lambda = \sqrt{\frac{2K}{\mu_\lambda}}\,, \hspace{0.3cm} \omega_{\rho_{cc}} = \sqrt{\frac{K}{\mu_{\rho_{cc}}}}\,, \label{eq:oscillator_frequcies_HO_Jacobian_coord} 
\end{align}
with the following ratios,
\begin{align}
    \frac{\omega_\lambda}{\omega_\rho} = \sqrt{\frac{1}{2}\left(1 + \frac{m_{u}}{m_{c}}\right)} \leq 1\,,
    \label{eq:lam/rho}
\end{align}
and 
\begin{align}
    \frac{\omega_\lambda}{\omega_{\rho_{cc}}} = \sqrt{\frac{1}{2}\left(1 + \frac{m_{c}}{m_{u}}\right)} \geq 1\,,
    \label{eq:lam/rhocc}
\end{align}
where $K$ denotes a spring constant and $\mu_i$ the reduced mass of the subsystems (see Eq.~\eqref{eq:reduced_mass_HO}).
As also shown in Fig.\@ \ref{fig:HO_energy_dep}, the mass hierarchy should undoubtedly be: $\omega_{\text{GS}} \leq \omega_{\rho_{cc}} \leq \omega_\lambda \leq \omega_\rho$. 
Besides, when the heavy-quark and light-quark masses are the same, i.e., $m_u = m_c$, the three excited modes are degenerated. However, for $m_c > m_u$, the $\rho$-mode energy is the highest excited energy of those.

Consequently, we would expect the same ordering in our $\overline{\rm LD}$HD approach based on the DQM, where the diquark masses are determined by solving the Schr\"{o}dinger equation with the Hamiltonian~(\ref{DQMHamiltonian}), but our numerically calculated $\rho$-mode energy lies between the $\rho_{cc}$-mode and the $\lambda$-mode.

To gain a deeper understanding of the inverted hierarchy, we calculate the centrifugal energy by taking the expectation value of the centrifugal potential
\begin{equation}
    \left\langle 
    \frac{l(l+1)}{2 \mu r^2}
    \right\rangle
    \ ,
\end{equation}
where $\mu$ is the relevant reduced mass. Since spatial size directly indicates internal structure and dynamics, we compute the Root-Mean-Square (RMS) radii 
\begin{align}
    \begin{split}
        r_\text{rms} =& \left(\int \text{d}^3r\, r^2|\psi(r)|^2\right)^{1/2}\,, \label{eq:RMS_integral}
    \end{split}
\end{align}
and expectation values of the individual Hamiltonian terms of the diquark subsystems and of the full tetraquark states, as summarized in Table \ref{tab:Tcc_Lamc_RMS_ExpVal}.
The RMS provides an intuitive measure of the compactness or extent of a state.
Changes in the RMS between ground and excited states reveal whether excitations are primarily radial, centrifugal (orbital), or driven by changes in effective constituent masses. 

This investigation provides an explanation for the counterintuitive inversion of the $\rho$- and $\lambda$-mode spectra. The effect originates from the nearly twofold larger value of $r_{\text{rms},ud}$ for the ($ud$) diquark compared with that of the $T_{cc}(0^-,1^-,2^-)_\lambda$ configuration. Although a smaller reduced mass $\mu_{ud}$ would naively suggest a larger centrifugal contribution $L_{T_{cc}(1^-;0^-,1^-,2^-;1^-,2^-,3^-)_\rho} = \langle l(l+1)/(2\mu_{ud} r^2)\rangle_{l=1}$ than for $L_{T_{cc}(0^-,1^-,2^-)_\lambda}$, the quadratic dependence on $r_{\text{rms},ud}$ enhances the sensitivity to the spatial size. Consequently, the larger radius dominates over the reduced-mass effect and ultimately determines the spectral ordering as $L_{T_{cc}(0^-,1^-,2^-)_\lambda}=0.481\text{~GeV}>L_{T_{cc}(1^-;0^-,1^-,2^-;1^-,2^-,3^-)_\rho} = 0.368\text{~GeV}$.

Below, we shall provide a detailed explanation of the mechanism that leads to the {\it inverse} hierarchy.
First, we should note that the RMS radius increases for states with higher orbital angular momentum, i.e., for the excited state and the $\lambda$-mode of the diquark and tetraquark, respectively. This broadening of the wave function naturally reflects the higher excitation energy, which requires a more spatially extended configuration.
In contrast, when increasing the ground-state mass of $T_{cc}$, namely by considering the $\rho_{(cc)}$-mode(s), where the diquark masses are increased, the energy levels rise, but the wave function corresponding to the relative motion between two diquarks becomes more compact. Thus, heavier diquark-diquark systems tend to have smaller RMS values, whereas higher angular-momentum states exhibit broader distributions. In other words, increasing the angular momentum extends the spatial separation between the two diquarks, whereas increasing their masses compresses it.

For the $\lambda$-mode, where the angular momentum is changed, we find that the expectation values of the kinetic and potential energy components combine to reproduce the total energy difference between the ground and excited states.
In contrast, for the $\rho_{(cc)}$-mode(s), both expectation values are small and do not sum to the observed energy gap. This behavior can be readily understood from Eq.\@ \eqref{eq:2pt_Hamiltonian_CoM_frame}: 
\begin{align}
    H = m_{cc} + m_{\bar{u}\bar{d}} + T + V\,.
\end{align}
In this case, the dominant contribution to the excitation energy arises not from the kinetic $T$ or potential terms $V$, but from the change in the diquark mass itself. The small residual variations in the expectation values stem from the dependence of both kinetic and potential terms on the reduced mass $\mu$, which shifts slightly when the diquark mass changes.

For clarifying the above observation that the change of the diquark mass $m_{ud}$ is essential for the mass difference between the $\rho$-mode excitation and the ground state, we change the mass of the excited light-diquark, $m_{ud}(0^-, 1^-, 2^-)$.
Interestingly, the naive ordering of the $T_{cc}$ excitations can be restored by adjusting $m_{ud}(0^-, 1^-, 2^-)$. We find a threshold value of $m_{ud}(0^-, 1^-, 2^-) = 1.225$ GeV, at which the two configurations become degenerate, $\omega_{\rho}^\text{thr} = \omega_\lambda$.
It should be noticed that both the $\lambda$-state and the $\rho$-threshold state exhibit nearly identical total excitation energies, yet the reduced masses and individual expectation values differ significantly, highlighting that the underlying excitation mechanism is distinct.
For the $\rho$-mode, the main contribution is given by the orbital angular momentum inside the light diquark, which contributes through the term $\langle l(l+1)/(2\mu_{ud} r^2)\rangle$.
Actually, a larger centrifugal energy can invert the ordering once more when the diquark mass is increased further, making the $\rho$-mode heavier than the $\lambda$-mode.
Hence, to restore the naive expectation we need the sufficient orbital excitation energy inside of the $m_{ud}(0^-, 1^-, 2^-)$ to exceed the one of $T_{cc}(0^-,1^-,2^-)_\lambda$, i.e., $L_{m_{ud}(0^-, 1^-, 2^-)}>L_{T_{cc}(0^-,1^-,2^-)_\lambda}$. 

Therefore, we can validate that the main source of the inversion is the centrifugal energy, since increasing the mass of the excited diquark manually adds more centrifugal energy to the system, and then leads to $L_{m_{ud}(0^-, 1^-, 2^-)}\geq 0.368$~GeV due to the compression of the wavefunction. Thus, if the ($\bar{u}\bar{d}$) diquark inside the $T_{cc}$ has enough mass, and hence enough angular momentum, the $\rho$-mode will be more energetic, i.e., $L_{m_{ud}(0^-, 1^-, 2^-)}>L_{T_{cc}(0^-,1^-,2^-)_\lambda} = 0.481$~GeV, restoring the naive ordering of the HO spectrum\footnote{How much the $L$- and $r_\text{rms}$-values actually change cannot be predicted with our method since adjusting parameters inside the $T_{cc}$ system leads to the loss of information of the underlying mechanism. Thus, by correcting the excited mass of the ($\bar{u}\bar{d}$) diquark, we can no longer look inside this subsystem and obtain all information.}.

\begin{figure}
    \centering
    \includegraphics[width=\linewidth]{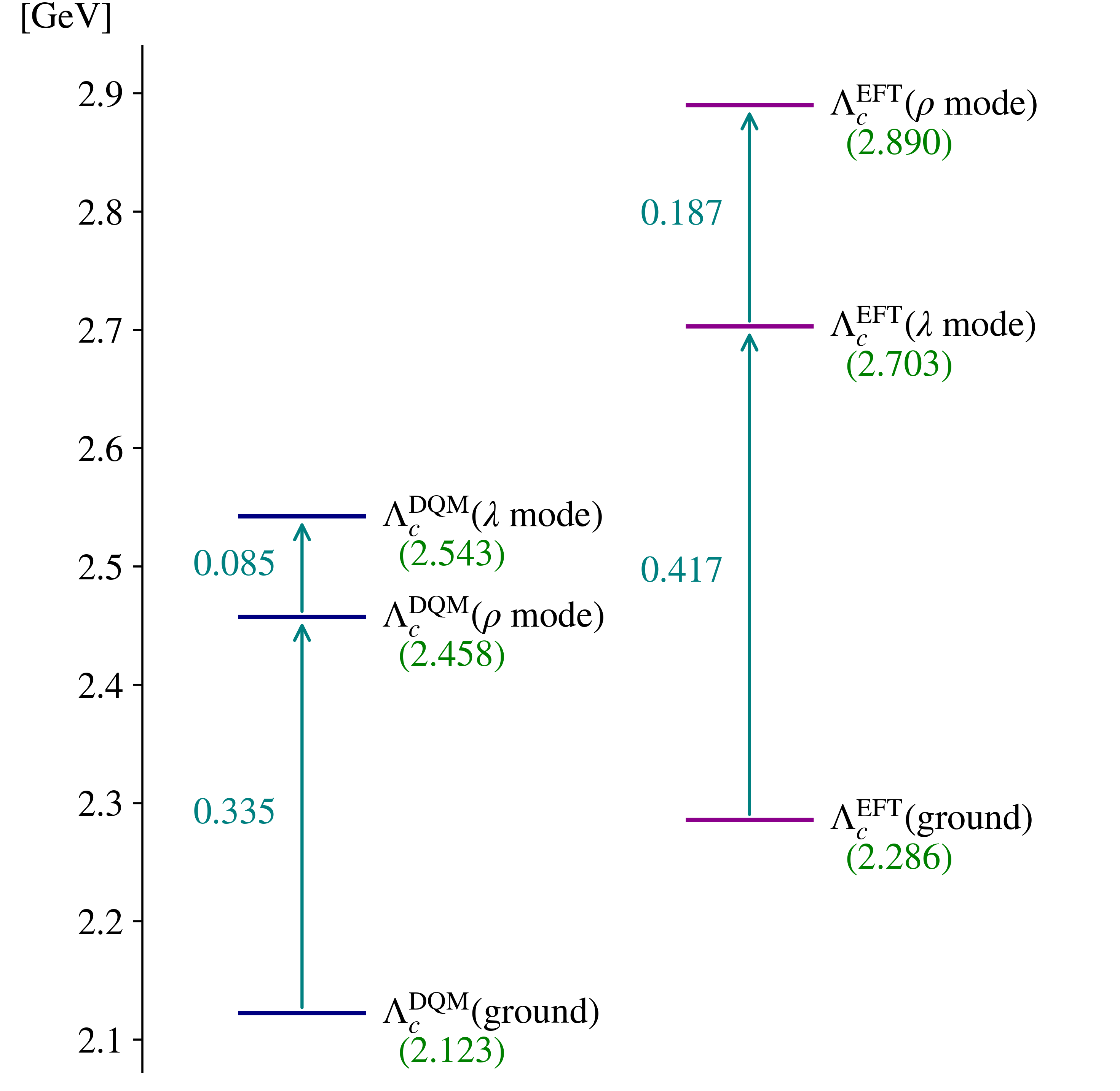}
    \caption{Comparison of the predicted $\Lambda_{c}$ spectrum obtained in the DQM (blue solid lines) with that from the chiral EFT framework \cite{PhysRevD.102.014004_Spectrum_of_singly_heavy_baryons__Kim} (violet solid lines). Same observation as for Fig.\@ \ref{fig:spectrum_Tcc} can be seen.}
    \label{fig:spectrum_Lamc_withdata}
\end{figure}

To confirm that this behavior is not unique to the $T_{cc}$ and to exclude any potential issues related to our choice of $m_{cc}$, we analyze the spectra of $\Lambda_c$ (see Fig.~\ref{fig:spectrum_Lamc_withdata}). 
Across these systems, our model consistently predicts the $\rho$-mode to lie below the $\lambda$-mode. This indicates that the mass gap between the ground and excited light-diquark states plays a decisive role in determining the relative ordering of these excitations. 

It should be noticed that a comparison between $T_{cc}$ and $\Lambda_c$ shows that, in the heavy-quark limit, both systems share the same static color configuration and thus obey the same superflavor symmetry~\cite{Georgi:1990ak,Savage:1990di,Tanaka:2024siw}.
Our results are consistent with this expectation: the energy difference between the ground state and the $\lambda$ excitation is almost the same, differing by $25$~MeV. 

As shown in Tab.~\ref{tab:Tcc_Lamc_RMS_ExpVal}, the RMS radius of (${u}{d}$) diquark is about $1.5$\,fm.
Now, in hadronic physics, a large RMS radius, typically above $\sim1$~fm, is often interpreted as indicative of a molecular configuration: two color-singlet hadrons weakly bound together, with most of the spatial extension arising from the inter-hadron separation. In such a picture, one can think of the system almost as a loosely bound molecule of hadrons.
Contrarily, in our $\overline{\rm LD}$HD description, the constituents are color-charged objects, e.g., a diquark and an anti-diquark. The system is not a color-singlet–color-singlet pair, but a compact multiquark bound state with internal color interactions. Therefore, even if the RMS radius appears large, it does not necessarily imply a molecular configuration, because the size now reflects the spatial distribution of colored objects rather than the separation of colorless hadrons.

It is worth emphasizing that in the GEM, the kinetic term includes both radial and angular contributions as shown in Eq.\@~\eqref{eq:Kinectic_term_GEM}.
With increasing angular momentum, the radial component broadens the wave function and effectively lowers the total energy, i.e., counteracting part of the angular contribution. This interplay explains why the overall kinetic expectation value is smaller than the pure angular term, demonstrating how the radial broadening stabilizes the system despite the centrifugal barrier.
This behavior can also be verified from the perspective of the uncertainty principle: $\Delta p \Delta x \geq \frac{\hbar}{2}$. As the radial wave function becomes more extended, the position uncertainty $\Delta x$ increases, implying a smaller momentum uncertainty $\Delta p$.
Following the calculation of 
\begin{align}
    \Delta\langle T\rangle = \left\langle \frac{\left(p_r^{(\text{ES})}\right)^2}{2\mu} \right\rangle - \left\langle \frac{\left(p_r^{(\text{GS})}\right)^2}{2\mu} \right\rangle + \left\langle \frac{l(l+1)}{2\mu r^2} \right\rangle_{l=1}\,,
\end{align}
we find that the average momentum in the excited state $p_r^{(\text{ES})}$ must indeed be smaller than in the ground state $p_r^{(\text{GS})}$, i.e., $p_r^{(\text{ES})}<p_r^{(\text{GS})}$.
Hence, the broadening of the wave function naturally leads to a reduction in the kinetic energy, consistent with the uncertainty principle.
\begin{table*}[]
    \centering
    \renewcommand{\arraystretch}{1.1}
    \setlength{\tabcolsep}{6pt}
    \begin{tabular}{l|cccccccc}
        \toprule[1.2pt]
        \toprule[1.2pt]
        State & $m_\text{ES}-m_\text{GS}$ & $\langle l(l+1)/(2\mu r^2)\rangle$ & $r_{\text{rms}}$ & $\mu$ & $\Delta\langle T\rangle$ & $\Delta\langle V_{\text{mag}}\rangle$ & $\Delta\langle V_{\text{Coul}}\rangle$ & $\Delta\langle V_{\text{lin}}\rangle$ \\
        \midrule[1.2pt]
        $m_{ud}(0^+)$ & 0.000 & 0.000 & 0.869 & 0.158 & $\phantom{+}0.000$ & $\phantom{+}0.000$ & $\phantom{+}0.000$ & $\phantom{+}0.000$ \\
        $m_{ud}(0^-,1^-,2^-)$ & 0.455 & 0.368 & 1.511 & 0.158 & $-0.041$ & $\phantom{+}0.182$ & $\phantom{+}0.047$ & $\phantom{+}0.267$ \\
        $m_{cc}(1^+)$ & 0.000 & 0.000 & 0.549 & 0.918 & $\phantom{+}0.000$ & $\phantom{+}0.000$ & $\phantom{+}0.000$ & $\phantom{+}0.000$ \\
        $m_{cc}(1^-)$ & 0.202 & 0.239 & 0.813 & 0.918 & $\phantom{+}0.039$ & $-0.010$ & $\phantom{+}0.060$ & $\phantom{+}0.113$ \\
        \midrule[0.8pt]
        $T_{cc}(1^+)$         & 0.000 & 0.000 & 0.495 & 0.559 & $\phantom{+}0.000$ & $\phantom{+}0.000$ & $\phantom{+}0.000$ & $\phantom{+}0.000$ \\
        $T_{cc}(0^-,1^-,2^-)_{\lambda}$   & 0.395 & 0.481 & 0.741 & 0.559 & $\phantom{+}0.067$ & $-0.020$ & $\phantom{+}0.137$ & $\phantom{+}0.210$ \\
        \makecell[l]{$T_{cc}(1^-;0^-,1^-,2^-;$\\ $\phantom{T_{cc}(}1^-,2^-,3^-)_\rho$}      & 0.313 & $(0.368)$ & 0.404 & 0.849 & $\phantom{+}0.003$ & $\phantom{+}0.000$ & $-0.075$ & $-0.069$ \\
        \makecell[l]{$T_{cc}^{\text{thr}}(1^-;0^-,1^-,2^-;$\\ $\phantom{T_{cc}^\text{thr}(}1^-,2^-,3^-)_\rho$} & 0.395 & $(\geq 0.368)$ & 0.391 & 0.907 & $\phantom{+}0.005$ & $\phantom{+}0.000$ & $-0.089$ & $-0.080$ \\
        \makecell[l]{$T_{cc}^\text{EFT}(1^-;0^-,1^-,2^-;$\\ $\phantom{T_{cc}^\text{EFT}(}1^-,2^-,3^-)_\rho$} & 0.606 & / & 0.363 & 1.042 & $\phantom{+}0.0010$ & $\phantom{+}0.000$ & $-0.122$ & $-0.101$ \\
        $T_{cc}(1^-)_{\rho_{cc}}$        & 0.199 & $(0.239)$ & 0.493 & 0.564 & $\phantom{+}0.000$ & $\phantom{+}0.000$ & $-0.001$ & $-0.002$ \\
        \midrule[0.8pt]
        $\Lambda_{c}(1/2^+)$     & 0.000 & 0.000 & 0.526 & 0.489 & $\phantom{+}0.000$ & $\phantom{+}0.000$ & $\phantom{+}0.000$ & $\phantom{+}0.000$ \\
        $\Lambda_{c}(1/2^-,3/2^-)_{\lambda}$ & 0.420 & 0.455 & 0.802 & 0.489 & $\phantom{+}0.052$ & $\phantom{+}0.000$ & $\phantom{+}0.132$ & $\phantom{+}0.236$ \\
        $\Lambda_{c}(1/2^-,3/2^-)_{\rho}$   & 0.335 & $(0.368)$ & 0.446 & 0.696 & $-0.003$ & $\phantom{+}0.000$ & $-0.056$ & $-0.061$ \\
        $\Lambda_{c}^{\text{thr}}(1/2^-,3/2^-)_{\rho}$ & 0.420 & $(\geq 0.368)$ & 0.435 & 0.734 & $-0.002$ & $\phantom{+}0.000$ & $-0.066$ & $-0.070$ \\
        $\Lambda_{c}^{\text{EFT}}(1/2^-,3/2^-)_{\rho}$ & 0.642 & / & 0.411 & 0.821 & $\phantom{+}0.000$ & $\phantom{+}0.000$ & $-0.087$ & $-0.088$ \\
        \bottomrule[1.2pt]
        \bottomrule[1.2pt]
    \end{tabular}
    \caption{Comparison of calculated properties for $T_{cc}$ and $\Lambda_c$ states. 
    All energies are given in GeV and $r_{\text{rms}}$ in fm, which describes the distance between (i) two quarks inside a diquark, (ii) two diquarks inside a tetraquark, and (iii) a diquark and quark inside a baryon. Values in parentheses denote the orbital excitation energy given by the diquark subsystem (whole system in ground state). Our model allows us to distinguish between all the different influences. The $\Delta$ indicates the difference between the excited-state (ES) and ground-state (GS) expectation value. The $V_\text{mag}$ corresponds only to the spin-spin interaction of Eq.~\eqref{SBPotential}.}
    \label{tab:Tcc_Lamc_RMS_ExpVal}
\end{table*}


\begin{figure}
  \centering
  \subfigure[]{
    \begin{tikzpicture}   
      \tikzset{
        particle/.style = {circle, draw, line width=1.8pt, inner sep=0pt, font=\Large},
        light/.style    = {particle, minimum size=1.4cm},
        heavy/.style    = {particle, minimum size=1.8cm}
      }
    
      \node[light] (Ltop)    at (0,  3) {$(\bar{u}\bar{d})_{0^-}$};
      \node[heavy] (Lbottom) at (0,  0) {$(cc)_{1^+}$};
    
      \node[light] (Rtop)    at (4,  3) {$(\bar{u}\bar{d})_{0^+}$};
      \node[heavy] (Rbottom) at (4,  0) {$(cc)_{1^+}$};
    
      \draw[line width=2pt]
        (Ltop) -- (Lbottom)
        node[midway, right, font=\large] {$L=0$};
      \draw[line width=2pt]
        (Rtop) -- (Rbottom)
        node[midway, right, font=\large] {$L=0$};
    
      \draw[->, line width=1pt, green!50!black!100]
        (Ltop) -- (Rtop)
        node[midway, above, font=\large, text=green!50!black!100] {$\eta_{I=0,L=0}$};
    
      \node[font=\Large, above=1.0cm of Ltop] {\(T_{cc}(1^-)_\rho\)};
      \node[font=\Large, above=1.0cm of Rtop] {\(T_{cc}(1^+)\)};

      \begin{scope}[on background layer]
        \node[draw, ellipse, dashed, line width=1.4pt, black!30, fit=(Ltop)(Lbottom), inner sep=1pt] (LeftEllipse) {};
        \node[draw, ellipse, dashed, line width=1.4pt, black!30, fit=(Rtop)(Rbottom), inner sep=1pt] (RightEllipse) {};
      \end{scope}


    \end{tikzpicture}
    \label{fig:eta_emission}
    }

  \vspace{2mm}

  \subfigure[]{
    \begin{tikzpicture}   
      \tikzset{
        particle/.style = {circle, draw, line width=1.8pt, inner sep=0pt, font=\Large},
        light/.style    = {particle, minimum size=1.4cm},
        heavy/.style    = {particle, minimum size=1.8cm}
      }
    
      \node[light] (Ltop)    at (0,  3) {$(\bar{u}\bar{d})_{0^+}$};
      \node[heavy] (Lbottom) at (0,  0) {$(cc)_{1^+}$};
    
      \node[light] (Rtop)    at (4,  3) {$(\bar{u}\bar{d})_{0^+}$};
      \node[heavy] (Rbottom) at (4,  0) {$(cc)_{1^+}$};
    
      \draw[line width=2pt]
        (Ltop) -- (Lbottom)
        node[midway, right, font=\large] {$L=1$};
      \draw[line width=2pt]
        (Rtop) -- (Rbottom)
        node[midway, right, font=\large] {$L=0$};
    
      \draw[->, line width=1pt, violet]
        (Ltop) -- (Rtop)
        node[midway, above, font=\large, text=violet] {$(\pi\pi)_{I=0,L=1}$};
    
      \node[font=\Large, above=1.0cm of Ltop] {\(T_{cc}(1^-)_\lambda\)};
      \node[font=\Large, above=1.0cm of Rtop] {\(T_{cc}(1^+)\)};

      \begin{scope}[on background layer]
        \node[draw, ellipse, dashed, line width=1.4pt, black!30, fit=(Ltop)(Lbottom), inner sep=1pt] (LeftEllipse) {};
        \node[draw, ellipse, dashed, line width=1.4pt, black!30, fit=(Rtop)(Rbottom), inner sep=1pt] (RightEllipse) {};
      \end{scope}


    \end{tikzpicture}
    \label{fig:pion_emission}
    }

  \caption{Illustration of \textbf{(a)} the $S$-wave $\eta$ (darkgreen) emission from the excited tetraquark $T_{cc}(1^-)_\rho$ and \textbf{(b)} the $P$-wave $\pi\pi$ (violet) emission from the excited tetraquark $T_{cc}(1^-)_\lambda$. The transition shows how the excited mode relaxes to the ground state by emitting an $\eta$ or two $\pi$ from the light-diquark, while the heavy ($cc$) diquark does not influence the decay.}
  \label{fig:eta_pion_emission}
\end{figure}
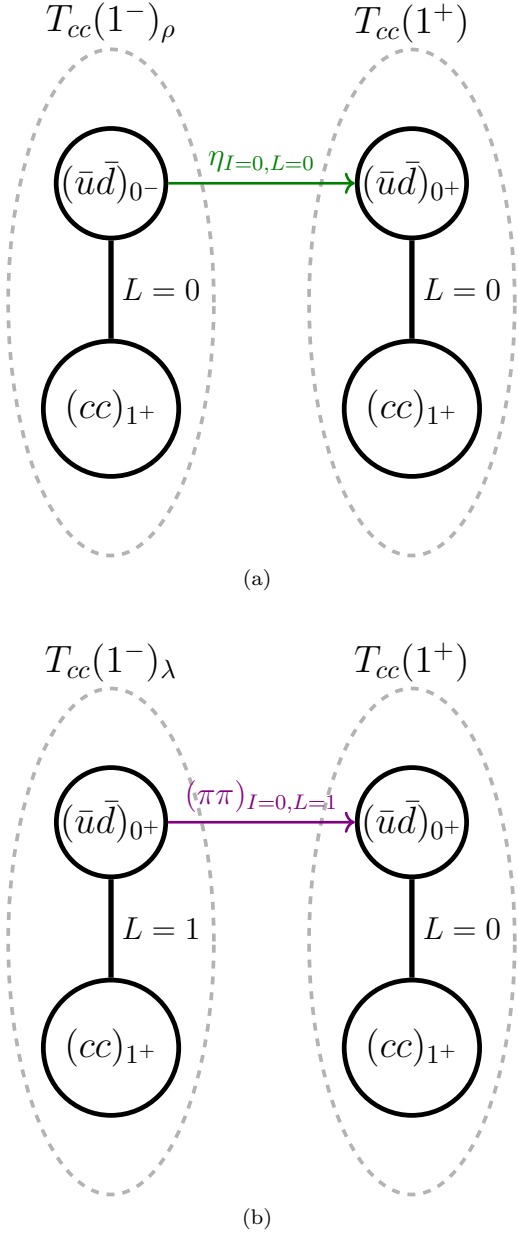

Before closing this section, we discuss how the $\rho_{cc}$-, $\rho$- and $\lambda$-mode excitations can be distinguished experimentally by focusing on their decay properties. First, we should note that the strongly decaying $\rho_{cc}$-mode is suppressed due to the heavy-quark spin symmetry (HQSS)~\cite{manohar2000heavy}: the charm diquark with $J^P = 1^-$ cannot flip into the one with $J^P = 1^+$ in the heavy quark limit.
Therefore, the ($cc$) diquark can be regarded as a spectator (and stable) in the heavy quark limit.
Hence, as illustrated in Fig.~\ref{fig:eta_pion_emission}, excited $T_{cc}$ states in $\lambda$- and $\rho$-modes may decay dominantly either through the emission of a single $\eta$ or two pions into the ground state.
If the excited configurations, corresponding to the $\lambda$- and $\rho$-modes, lie above the $\eta$-emission threshold, one expects, e.g., the following decay channels:
\begin{align}
    \begin{split}
        T_{cc}(1^-)_\rho &\xrightarrow[]{\phantom{.(}\eta_{I=0,L=0}\phantom{.)}} T_{cc}(1^+)\,,\\
        T_{cc}(1^-)_\lambda &\xrightarrow[]{(\pi\pi)_{I=0, L=1}} T_{cc}(1^+)\,,
    \end{split} \label{DecayMode}
\end{align}
where a dominant $S$-wave ($L=0$) $\eta$ decay (see Fig.~\ref{fig:eta_emission}) would indicate the $\rho$-mode, whereas either a significant $P$-wave ($L=1$) two-pion (see Fig.~\ref{fig:pion_emission}) signal or a suppressed $\eta$ signal would be characteristic of the $\lambda$-mode.
Conversely, if the $\eta$ threshold lies above both excited states, only two-pion transitions are allowed. In this case, a prominent $\pi\pi$ decay would be a signature of the $\lambda$-mode. 



The above selection rules are not expected to be exact in realistic systems due to the small violation of HQSS. 
But still the main decay modes follow Eq.~\eqref{DecayMode}. Therefore, if the excited $T_{cc} (1^-)$ state will be observed above the $\eta$ threshold, the substantial $\eta$ decay and $\pi\pi$ decay imply its $\lambda$-mode and $\rho$-mode dominant nature, respectively. Meanwhile, if it lies below the threshold, then the significant $\pi\pi$ decay corresponds to its $\lambda$-mode dominant nature while the comparably suppressed $\pi\pi$ decay indicates its $\rho$-mode one. 

\section{Additional derivations for \texorpdfstring{$T_{bb}$ and $\Lambda_b$}{Tbb and Lambdab}} \label{sec:Tbb_and_Lamb}
    \begin{figure}[!b]
    \centering
    \includegraphics[width=\linewidth]{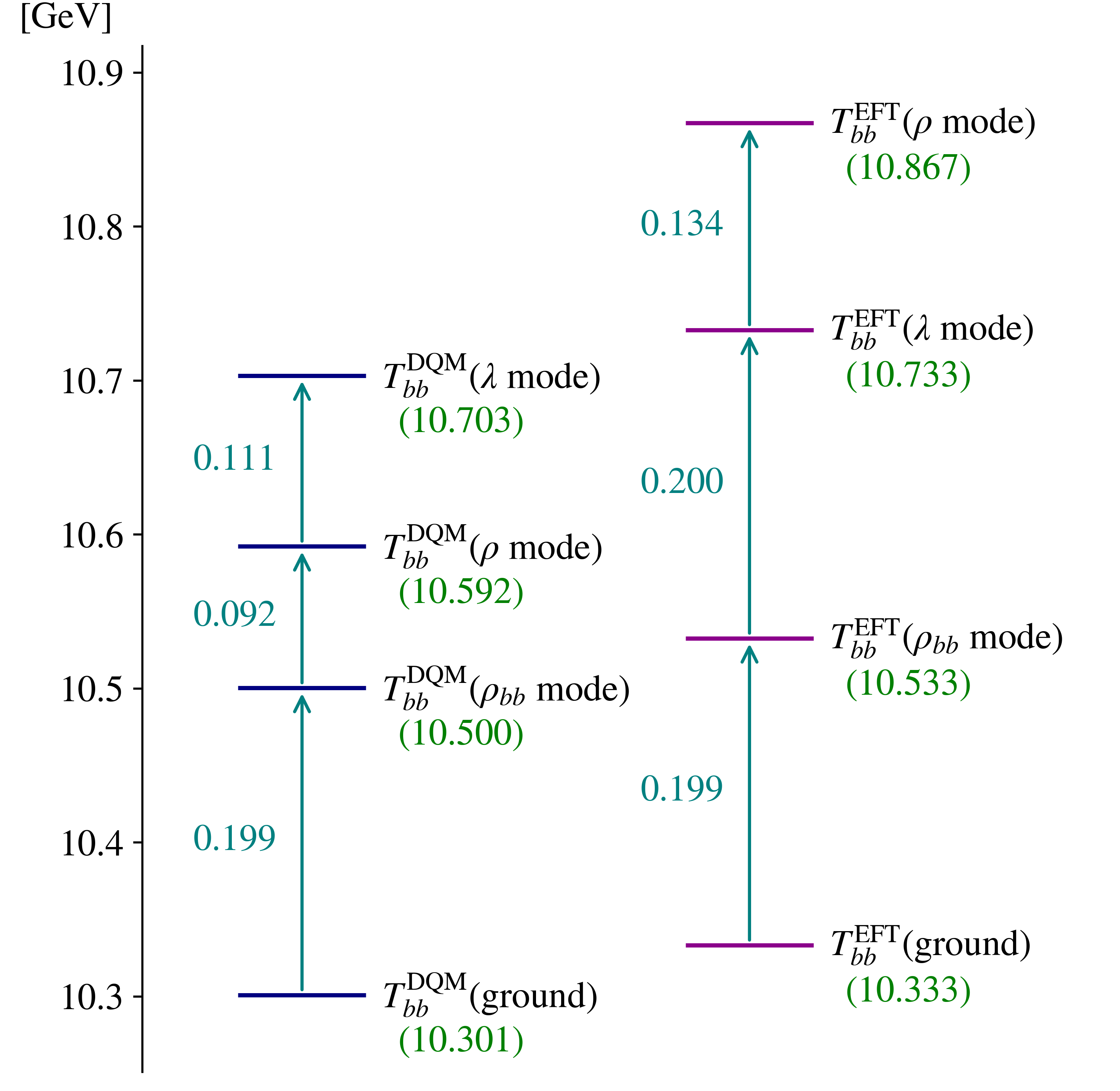}
    \caption{Same as in Fig.~\ref{fig:spectrum_Tcc} but for $T_{bb}$.}
    \label{fig:Tbb_spectrum}
\end{figure}

To verify that the observed mass hierarchy between the $\rho$- and $\lambda$-modes is not restricted to the charm sector, we repeat the same analysis for the bottom counterparts $T_{bb}$ and $\Lambda_b$. 
The formalism and parameter setup are identical to those described in the main text for the $T_{cc}$ and $\Lambda_c$ systems, except for the replacement of charm-quark parameters by bottom-quark ones.

\begin{figure}
    \centering
    \includegraphics[width=\linewidth]{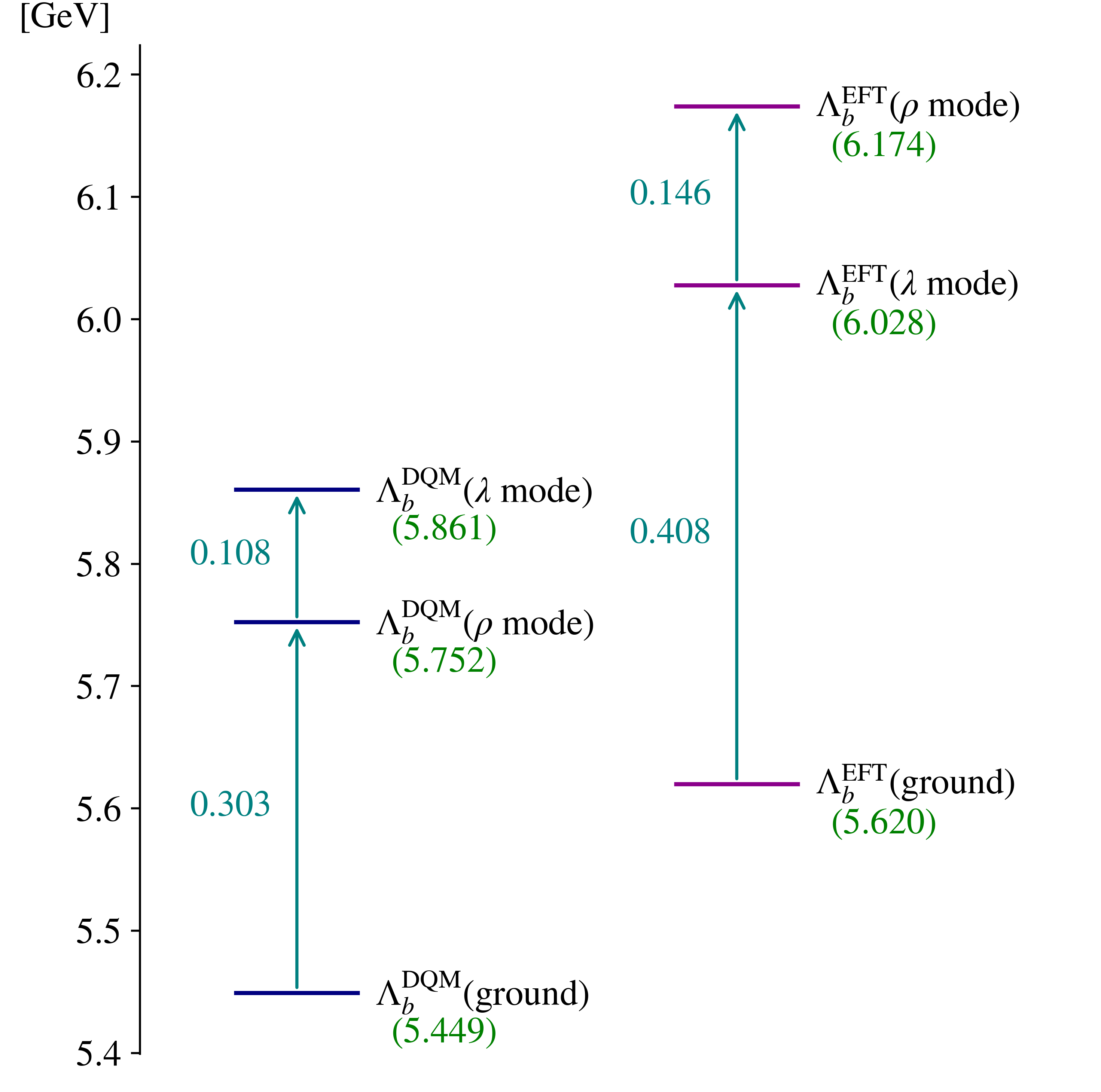}
    \caption{Same as in Fig.~\ref{fig:spectrum_Lamc_withdata} but for $\Lambda_b$.}
    \label{fig:Lamb_spectrum}
\end{figure}

As shown in Fig.~\ref{fig:Tbb_spectrum}, Fig.~\ref{fig:Lamb_spectrum} and Tab.~\ref{tab:Tbb_Lamb}, 
the same qualitative behavior is found: the $\rho$-mode lies below the {$\lambda$-mode}, contrary to the naive expectation from an HO picture. 
This demonstrates that the inverted hierarchy originates primarily from the light-diquark dynamics rather than from heavy-quark specific dynamics.

The consistency between charm and bottom sectors supports our interpretation that the crucial factor controlling the mode ordering is the mass gap between the light-diquark ground and excited states, i.e., the centrifugal energy between two light quarks.
Thus, even though the absolute energy scales differ significantly between the two sectors, the underlying mechanism remains unchanged.

A comparison of $T_{bb}$ and $\Lambda_b$ in the heavy-quark limit leads to the same conclusion as in the charm sector. 
This is the reflection of the heavy-flavor symmetry, i.e., excitation energies of the charm sector are identical to those of the bottom sector in the heavy quark limit.
In all cases, the relative differences are identical, providing further evidence for the internal consistency of our calculations.

\begin{table*}[]
    \centering
    \renewcommand{\arraystretch}{1.1}
    \setlength{\tabcolsep}{6pt}
    \begin{tabular}{l|cccccccc}
        \toprule[1.2pt]
        \toprule[1.2pt]
        State & $m_\text{ES}-m_\text{GS}$ & $\langle l(l+1)/(2\mu r^2 )\rangle$ & $r_\text{rms}$ & $\mu$ & $\Delta \langle T\rangle$ & $\Delta \langle V_{\text{mag}}\rangle$ & $\Delta \langle V_{\text{Coul}}\rangle$ & $\Delta \langle V_{\text{lin}}\rangle$ \\
        \midrule[1.2pt]
        $m_{bb}(1^+)$ & 0.000 & 0.000 & 0.322 & 2.614 & $\phantom{+}0.000$ & $\phantom{+}0.000$ & $\phantom{+}0.000$ & $\phantom{+}0.000$ \\
        $m_{bb}(1^-)$ & 0.200 & 0.198 & 0.537 & 2.614 & $-0.014$ & $-0.004$ & $\phantom{+}0.127$ & $\phantom{+}0.091$ \\
        \midrule[0.8pt]
        $T_{bb}(1^+)$        & 0.000 & 0.000 & 0.470 & 0.625 &  $\phantom{+}0.000$ &  $\phantom{+}0.000$ &  $\phantom{+}0.000$  &  $\phantom{+}0.000$ \\
        $T_{bb}(0^-,1^-,2^-)_{\lambda}$  & 0.402 & 0.447 & 0.722 & 0.625 &  $\phantom{+}0.043$ & $-0.007$ &  $\phantom{+}0.151$  &  $\phantom{+}0.215$ \\
        \makecell[l]{$T_{bb}(1^-;0^-,1^-,2^-;$\\ $\phantom{T_{cc}(}1^-,2^-,3^-)_\rho$}     & 0.291 & $(0.368)$ & 0.370 & 1.009 &  $\phantom{+}0.010$ &  $\phantom{+}0.000$ & $-0.096$  & $-0.077$ \\
        \makecell[l]{$T_{bb}^\text{thr}(1^-;0^-,1^-,2^-;$\\ $\phantom{T_{cc}^\text{thr}(}1^-,2^-,3^-)_\rho$} & 0.402 & $(\geq0.368)$ & 0.348 & 1.129 &  $\phantom{+}0.014$ &  $\phantom{+}0.000$ & $-0.116$  & $-0.089$ \\
        \makecell[l]{$T_{bb}^\text{EFT}(1^-;0^-,1^-,2^-;$\\ $\phantom{T_{bb}^\text{EFT}(}1^-,2^-,3^-)_\rho$} & 0.566 & / & 0.322 & 1.294 &  $\phantom{+}0.025$ &  $\phantom{+}0.000$ & $-0.164$  & $-0.113$ \\
        $T_{bb}(1^-)_{\rho_{bb}}$   & 0.199 & $(0.198)$ & 0.470 & 0.625 & $\phantom{+}0.000$ & $\phantom{+}0.000$ & $\phantom{+}0.000$  & $\phantom{+}0.000$ \\
        \midrule[0.8pt]
        $\Lambda_{b}(1/2^+)$        & 0.000 & 0.000 & 0.495 & 0.559 &  $\phantom{+}0.000$ &  $\phantom{+}0.000$ &  $\phantom{+}0.000$  &  $\phantom{+}0.000$ \\
        $\Lambda_{b}(1/2^-,3/2^-)_{\lambda}$  & 0.411 & 0.438 & 0.746 & 0.591 &  $\phantom{+}0.038$ &  $\phantom{+}0.000$ &  $\phantom{+}0.148$  &  $\phantom{+}0.225$ \\
        $\Lambda_{b}(1/2^-,3/2^-)_{\rho}$     & 0.303 & $(0.368)$ & 0.387 & 0.923 &  $\phantom{+}0.006$ &  $\phantom{+}0.000$ & $-0.084$  & $-0.073$ \\
        $\Lambda_{b}^{\text{thr}}(1/2^-,3/2^-)_{\rho}$ & 0.411 & $(\geq0.368)$ & 0.368 & 1.017 &  $\phantom{+}0.010$ &  $\phantom{+}0.000$ & $-0.107$  & $-0.088$ \\
        $\Lambda_{b}^{\text{EFT}}(1/2^-,3/2^-)_{\rho}$ & 0.588 & / & 0.343 & 1.156 &  $\phantom{+}0.017$ &  $\phantom{+}0.000$ & $-0.141$  & $-0.107$ \\
        \bottomrule[1.2pt]
        \bottomrule[1.2pt]
    \end{tabular}
    \caption{Comparison of the $T_{bb}$ and $\Lambda_b$ systems. Shown are mass differences, kinetic and potential energy contributions, and size parameters.}
    \label{tab:Tbb_Lamb}
\end{table*}

\section{Conclusion} \label{sec:Conclusion}
    In this work, we have shown that applying the diquark model to the $T_{cc}$ system yields an unexpected mass hierarchy. Contrary to the naive expectation from an HO picture, in which the $\rho$-mode is energetically higher than the $\lambda$-mode, our results place the $\rho$-mode below the latter.
We conclude that the main source of the inversion originates from the centrifugal energies.
The centrifugal energy for $\rho$-mode is given by the relative motion between $u$ and $d$ quarks inside ($ud$) diquark, while for $\lambda$-mode it is by the one between ($ud$) diquark and ($cc$) diquark.
Although the reduced mass of $u$ and $d$ quarks is much smaller than that of the ($ud$) and ($cc$) diquarks, the RMS distance between $u$ and $d$ quarks compensate that between the two diquarks.
As a result, the centrifugal energy for $\rho$-mode is smaller than that for $\lambda$-mode, which provided the inverted hierarchy.
To further verify this behavior, we studied the spectra, the RMS radii, and expectation values for several systems ($T_{cc}$, $T_{bb}$, $\Lambda_c$, $\Lambda_b$). We find that the inverted hierarchy between the $\rho$ and $\lambda$ modes occurs in all systems.
Thus, we found that the $\rho$-mode might be lighter than the $\lambda$-mode and that the experimentally measured excited $\lambda$-states might be the $\rho$ ones. 
Moreover, if the threshold is not opened, the $\rho$- or $\lambda$-mode can not be distinguished, and hence $\rho$-mode might be lighter than $\lambda$-mode.

What does this inverse ordering now concretely imply for our current understanding? 
Such an analysis provides an intuitive connection between the hadronic excitation spectrum and the underlying chiral dynamics. 
Since the $\rho$-mode excitation with $J^P=1^-$ will be the chiral partner to the ground state, it is important to distinguish the $\rho$-mode from the $\lambda$-mode for obtaining some information on the chiral dynamics in the heavy hadron sector.
Our result in this paper implies that we may not be able to identify the $\rho$-mode just from the mass spectrum.
Then, for experimentally distinguishing the $\rho$-mode from the $\lambda$-mode, we provided a discussion on the decay modes of the excitations: 
It is essential to compute the decay of the $\rho$-mode and compare it to the $\eta$ or $\pi\pi$ decay, since the chiral partner couples to the ground state through this distinct emission.
A further promising direction for future work is to investigate the dependence of the spectra on the choice of the different potential models, such as that of T.~Yoshida \textit{et al.\@} \cite{PhysRevD.92.114029_Spectrum_of_heavy_baryons__Yoshida}, which distinguishes between different isospin channels, or the model by T.~Barnes \textit{et al.\@} \cite{PhysRevD.72.054026_Higher_charmonia__Barnes}, which uses a mass-dependent Coulomb term motivated by lattice QCD results. 
However, since the former potential includes $L\cdot S$ terms, we expect that our degenerate states will be separated, but only the $2^-$ state will experience an energy shift upward. The other might be corrected downward. Hence, the $m_{ud}(0^-)$ and $m_{ud}(1^-)$ will be below our predicted energy, and consequently, the corresponding $T_{cc}$ states will also be corrected.

An interesting extension of the present work would be the investigation of the $\Xi_{c}$ and $\Xi_{b}$ mass spectra within the same framework. In particular, systems containing a $us$ or $ds$ diquark may exhibit different dynamical behavior due to the larger strange-quark mass. Since these diquark masses are expected to lie closer to the corresponding excitation ``thresholds"\footnote{``Threshold" refers here to the point at which the $\rho$-mode got heavier than the $\lambda$-mode, see Sec.~\ref{sec:results}.}, it would be worthwhile to study how the structure and excitation properties of such hadrons are affected in comparison to the non-strange sector. 
However, a preliminary study of the $\Xi_c$ and $\Xi_b$ systems within the present framework suggests a behavior similar to that observed here. Since the primary effect of replacing a light quark by a strange quark is an increase in the diquark mass, the same qualitative arguments used in the comparison between the $T_{cc}$ and $\Lambda_c$ systems appear to remain valid for the $\Xi_c$ baryon, and thus $\Xi_b$, respectively. In this picture, the larger strange-quark mass mainly shifts the relevant mass scales and excitation energies without modifying the underlying dynamics responsible for the interplay between the $\rho$- and $\lambda$-mode excitations significantly. 
Nevertheless, this observation should be confirmed through a dedicated investigation. Such a study would be particularly interesting when combined with an analysis of the flavor partner of the $T_{cc}$, namely the $T_{ccs}$, as it could provide further insight into the role of strange quarks and the applicability of the diquark picture in heavy-hadron spectroscopy.

An additional natural extension would be to treat the system as a three-body configuration rather than a two-body approximation, thereby including correlations beyond the simple diquark picture. This direction will be pursued in future work.
It is expected that those studies, including the investigation of decay properties, will provide future experiments with useful information on light-quark dynamics in the exotic $T_{cc}$.



Another possible way to verify whether this unexpected level inversion is indeed physical would be to employ an interaction potential derived from lattice QCD within the HAL QCD framework, where the underlying quark–diquark interaction may be described more reliably. Therefore, it would be worthwhile to follow the approach of Ref.~\cite{Kelvin_Lee:20250c} and investigate the spectrum using the corresponding fitted potential.

\begin{acknowledgments}
    The work of M.~W.\@ is financially supported by JST SPRING, Grant Number JPMJSP2125. M.~W.\@ would also like to use the chance to thank the ``Tokai Higher Education and Research System Make New Standards Program for the Next Generation Researchers".
    The work of D.~S.\@ is supported in part by Grants-in-Aid for Scientific Research No.~23K03377, No.~23H05439, and No.~25K17386.
    The work of M.~H.\@ is supported in part by Grants-in-Aid for Scientific Research No.~23H05439 and No.~24K07045.
\end{acknowledgments}

\appendix
\section{Gaussian Expansion Method} \label{app:GEM}
    The GEM was proposed in 1988 by Kamimura \cite{PhysRevA.38.621_Nonadiabatic__Kamimura}, and is used to solve bound and scattering states for few-body systems. 
Since then it has found a broad application solving those systems.
In order to accurately describe long-range asymptotic behavior, short-range correlations, and the highly oscillatory nature of wave functions in the bound and scattering states of the systems, a carefully selected set of Gaussian basis functions forms an approximate complete set in a finite coordinate space. The foundations of the GEM are that
\begin{enumerate}[label=(\roman*)]
    \item The basis set consists of functions of the Jacobian coordinates, and
    \item The radial dependence is Gaussian $\phi_{nl} = r^l e^{-\nu_n r^2}$, with the range parameters forming a geometric progression, $\left\{\nu_n = \nu_1 a^{n-1}; n = 1, \dots, n_{\text{max}} \right\}$. \label{item:GEM_radial_dep}
\end{enumerate}
Due to property \ref{item:GEM_radial_dep} for the Gaussian ranges, the short-range correlations as well as the long-range tail behavior in the asymptotic region are captured precisely.
Moreover, the advantage of this specific shape is that matrix elements between basis functions of different channels can be obtained effortlessly \cite{HIYAMA2003223_GEM__Hiyama, 10.1093/ptep/pts015_GEM__Hiyama, GEM__Hiyama}.

For our case, we apply the GEM to a two-body system and, therefore, briefly explain the method for this special case. 

In GEM, the solution to the Schr\"odinger equation for the bound-state wave function $\psi_{JM}$ of a few-body system with the total angular momentum $J$ and its $z$-component $M$,
\begin{align}
    (H - E)\psi_{JM} = 0\,,
\end{align}
is solved by using the Rayleigh-Ritz variational method. For simplicity, other quantum numbers such as parity and isospin are neglected. 
Next, we expand the total wave function $\psi_{JM}$ in terms of a set of $L^2$-integrable basis functions $\left\{\phi_{JM,n}; n = 1,\dots,n_{\text{max}}\right\}$ as
\begin{align}
    \psi_{JM} = \sum_n C_n^{(J)} \phi_{JM,n}\,,
\end{align}
where the eigenenergy $E$ and the coefficients $C_n^{(J)}$ are given by
\begin{align}
    \langle \phi_{JM,n} | H - E | \psi_{JM} \rangle = 0\,.
\end{align}
Eventually, the Rayleigh-Ritz principle leads to a generalized matrix eigenvalue problem
\begin{align}
    \sum_{n' = 1}^{n_\text{max}} \left(H_{nn'}^{(J)} - EN_{nn'}^{(J)} \right) C_{n'}^{(J)} = 0\,, \label{eq:Rayleigh-Ritz_variational_principle}
\end{align}
with the following energy and overlap matrix elements:
\begin{align}
    \begin{alignedat}{2}
        H_{nn'}^{(J)} &= \langle \phi_{JM,n} | &H& | \phi_{JM,n'} \rangle \,, \\
        N_{nn'}^{(J)} &= \langle \phi_{JM,n} | &1& | \phi_{JM,n'} \rangle \,.
  \end{alignedat}
\end{align}
Hence, the lowest and excited state eigenfunctions with the same $J$ and parity can both be obtained by solving the eigenvalue problem \cite{10.1093/ptep/pts015_GEM__Hiyama, HIYAMA2003223_GEM__Hiyama, GEM__Hiyama}. 

Let us now specifically use the two-body Schr\"odinger equation
\begin{align}
    \left[ -\frac{\hbar^2}{2\mu}\nabla^2 + V(r) - E \right]\psi_{lm}(\mathbf{r}) = 0\,,
\end{align}
where $\mu$ is the reduced mass and $V(r)$ an arbitrary potential, to make this method more comprehensible.
By expanding $\psi_{lm}(\mathbf{r})$ in terms of the previously proposed Gaussian basis functions, we obtain 
\begin{align}
    &\psi_{lm}(\mathbf{r}) = \sum_n c_{nl} \phi_{nlm}^G(\mathbf{r})\,,\\
    &\phi_{nlm}^G(\mathbf{r}) = \phi_{nl}^G(r) Y_{lm}(\hat{\mathbf{r}})\,,\\
    &\phi_{nl}^G(r) = N_{nl}\ r^l e^{-\nu_n r^2}\,,\\
    &N_{nl} = \left( \frac{2^{l+2}(2\nu_n)^{l+\frac{3}{2}}}{\sqrt{\pi}(2l+1)!!}\right)^{\frac{1}{2}}\,,
\end{align}
where $Y_{lm}(\hat{\mathbf{r}})$ (or $Y_{lm}(\theta, \varphi)$) are the Laplace's spherical harmonics, and the constant $N_{nl}$ if for normalizing $\langle \phi_{nlm}^G | \phi_{nlm}^G \rangle = 1$, since the set $\left\{\phi_{nlm}^G; n=1,\dots,n_\text{max} \right\}$ is a non-orthogonal set. 
The best set of Gaussian size parameters is those in geometric progression
\begin{align}
    \nu_n &= r_n^{-2}\,, \\
    r_n &= r_1 a^{n-1}\,.
\end{align}
The basis functions' spatial ranges extend from very compact to very diffuse, with a denser coverage in the short-distance region than at long distances. The short-range basis functions (with small range parameters) accurately describe the fine, short-range structure of the system, while the long-range basis functions (with large range parameters) capture the correct asymptotic behavior of the wavefunction. 

By applying the Rayleigh-Ritz variational principle, the expansion coefficients $c_{n'l}$ and the eigenenergy $E$ are defined by the generalized matrix eigenvalue problem
\begin{align}
    \sum_{n'=1}^{n_\text{max}} \left[ \left( T_{nn'} + V_{nn'} \right) - EN_{nn'} \right] c_{n'l} = 0\,, \label{eq:generalized_EV_problem}
\end{align}
with the matrix elements \cite{HIYAMA2003223_GEM__Hiyama, 10.1093/ptep/pts015_GEM__Hiyama}
\begin{align}
    N_{nn'} &= \langle \phi_{nlm}^G | \phi_{n'lm}^G \rangle = \left( \frac{2\sqrt{\nu_n \nu_{n'}}}{\nu_n + \nu_{n'}} \right)^{l + \frac{3}{2}}\,, \\
    \begin{split}
       T_{nn'} &= \langle \phi_{nlm}^G | -\frac{\hbar^2}{2\mu}\nabla^2 | \phi_{nlm}^G \rangle  \\ &= \frac{\hbar^2}{\mu} \frac{(2l+3)\nu_n \nu_{n'}}{\nu_n + \nu_{n'}} \left( \frac{2\sqrt{\nu_n \nu_{n'}}}{\nu_n + \nu_{n'}}\right)^{l + \frac{3}{2}}\,, \label{eq:Kinectic_term_GEM}
    \end{split} \\
    \begin{split}
        V_{nn'} &= \langle \phi_{nlm}^G | V(r) | \phi_{n'lm}^G \rangle \\
        &= N_{nl} N_{n'l} \int_0^\infty \text{d}r\ r^{2(l+1)} e^{-(\nu_n + \nu_{n'})r^2} V(r) \,. \label{eq:GEM_potential_matrix_form}
    \end{split}
\end{align}
Hence, for the RMS radius, we obtain the following:
\begin{align}
    \begin{split}
        r_\text{rms} =& \left(\int \text{d}^3r\, r^2|\psi(r)|^2\right)^{1/2} \\
        =& \sum_{n,n'}\left( c_{nl} c_{n'l} \frac{l + \frac{3}{2}}{\nu_n + \nu_{n'}} \left( \frac{2\sqrt{\nu_n \nu_{n'}}}{\nu_n + \nu_{n'}} \right)^{l+\frac{3}{2}} \right)^{1/2}\,, \label{eq:RMS}
    \end{split}
\end{align}
where we used
\begin{align}
    \langle \phi_{nlm}^G | r^2 | \phi_{n'lm}^G \rangle = \frac{l+\frac{3}{2}}{\nu_n + \nu_{n'}}\left( \frac{2\sqrt{\nu_n \nu_{n'}}}{\nu_n + \nu_{n'}} \right)^{l + \frac{3}{2}}\,. \label{eq:r2_GEM_helper}
\end{align}

\section{Harmonic Oscillator Model} \label{app:HO}
    In the HO model, hadrons are treated as bound systems of constituent quarks moving in an effective confining potential, where the leading-order approximation is quadratic in the inter-quark separation. 
The corresponding Hamiltonian in its general form, 
\begin{align}
    H = \sum_i \frac{\mathbf{p}_i^2}{2m_i} + \frac{1}{2}K \sum_{i<j} \mathbf{r}_{ij}^2\,, \label{eq:HO}
\end{align}
with the momentum $\mathbf{p}_i$ of a quark $i$ with related mass $m_i$, an experimentally calculated spring constant $K$, and the inter-quark separation $\mathbf{r}_{ij}$, yields analytically solvable wave functions and a regular spacing of excitation energies \cite{PhysRevD.107.034031_Strong_decay_widths__Tecocoatzi, PhysRevD.73.114011_Covariant_oscillator_quark_model__Buisseret}. 
Hence, the eigenvalues scale according to
\begin{align}
    E_n \propto \left(2n + l + \frac{3}{2}\right)\hbar\omega\,,
\end{align}
where $l$ denotes the orbital angular momentum.
Consequently, the naive mass hierarchy ordering appears naturally from the dependence of the oscillator frequency $\omega$ on the reduced mass:
\begin{align}
    \omega\sim \sqrt{\frac{K}{\mu}}\,,
\end{align}
in such a manner that subsystems with smaller reduced mass $\mu$ yield larger excitation energies. Thus, since the light-light ($\rho$) reduced mass is smaller than the heavy-light ($\lambda$) reduced mass, and thus its oscillator frequency is larger, one expects $\rho$-mode excitations to lie above the $\lambda$-excitations. This ordering -- $\lambda$ below $\rho$ -- has become the standard expectation in heavy baryons \cite{PhysRevD.107.034031_Strong_decay_widths__Tecocoatzi, PhysRevD.73.114011_Covariant_oscillator_quark_model__Buisseret}. 

Although this simple oscillator picture omits important dynamical effects such as spin-dependent fores, relativistic corrections, and the true from of the QCD potential, it serves as a valuable benchmark for spectral ordering \cite{PhysRevD.107.034031_Strong_decay_widths__Tecocoatzi}. 
Deviations from the naive HO hierarchy in relativistic quark models therefore provide direct insight into the underlying QCD dynamics, i.e., the role of the centrifugal forces, spin-orbit couplings, and diquark correlations.

Let us briefly show how our naive intuition can be verified. As above truncated, the baryon system can be explained via using a nonrelativistic quark model with a HO potential for confinement.
By introducing the Jacobian coordinates $\boldsymbol{\lambda} = \frac{\mathbf{r}_{c_1} + \mathbf{r}_{c_2}}{2} - \frac{\mathbf{r}_u + \mathbf{r}_d}{2}$, $\boldsymbol{\rho} = \mathbf{r}_u - \mathbf{r}_d$, and $\boldsymbol{\rho_{cc}} = \mathbf{r}_{c_1} - \mathbf{r}_{c_2}$, with an intuitive notation, the HO Hamiltonian \eqref{eq:HO} takes the form
\begin{align}
    \begin{split}
        H =& \frac{\mathbf{p}_\rho^2}{2\mu_\rho} + \frac{\mathbf{p}_\lambda^2}{2\mu_\rho} + \frac{\mathbf{p}_{\rho_{cc}}^2}{2\mu_{\rho_{cc}}} \\ 
        &+ \frac{\mu_\rho\omega_\rho^2}{2}\boldsymbol{\rho}^2 + \frac{\mu_\lambda\omega_\rho^2}{2}\boldsymbol{\lambda}^2 + \frac{\mu_{\rho_{cc}}\omega_{\rho_{cc}}^2}{2}\boldsymbol{\rho}_{cc}^2\,, \label{eq:HO_Jacobian_coord}
    \end{split}
\end{align}
where $\mu_i$ denotes the reduced mass of $i\in\{\rho, \lambda, \rho_{cc}\}$,
\begin{align}
    \mu_{\rho}=\frac{m_u}{2}\,, \hspace{0.3cm} \mu_{\lambda}=\frac{2m_{u}m_{c}}{m_{u} + m_{c}}\,, \hspace{0.3cm} \mu_{\rho_{cc}}=\frac{m_{c}}{2}\,. \label{eq:reduced_mass_HO}
\end{align}
In addition, the oscillator frequencies are evaluated by
\begin{align}
    \omega_\rho = \sqrt{\frac{K}{\mu_\rho}}\,, \hspace{0.3cm} \omega_\lambda = \sqrt{\frac{2K}{\mu_\lambda}}\,, \hspace{0.3cm} \omega_{\rho_{cc}} = \sqrt{\frac{K}{\mu_{\rho_{cc}}}}\,, 
    \tag*{\eqref{eq:oscillator_frequcies_HO_Jacobian_coord}}
\end{align}
with the following ratios,
\begin{align}
    \frac{\omega_\lambda}{\omega_\rho} = \sqrt{\frac{1}{2} \left(1 + \frac{m_{u}}{m_{c}}\right)} \leq 1\,,
    \tag*{\eqref{eq:lam/rho}}
\end{align}
and 
\begin{align}
    \frac{\omega_\lambda}{\omega_{\rho_{cc}}} = \sqrt{\frac{1}{2} \left(1 + \frac{m_{c}}{m_{u}}\right)} \geq 1\,,
    \tag*{\eqref{eq:lam/rhocc}}
\end{align}
As shown in Fig.~\ref{fig:HO_energy_dep}, the mass hierarchy should be: $\omega_{\text{GS}} \leq \omega_{\rho_{cc}} \leq \omega_\lambda \leq \omega_\rho$. 
When the heavy-quark and light-quark masses are the same, i.e., $m_u = m_c$, the three excited modes are degenerated. But, for $m_c > m_u$, the $\rho$-mode energy is the highest excited energy of those.

\bibliographystyle{apsrev4-2}
\bibliography{bibliography}

\end{document}